\documentclass[prd,aps,nofootinbib]{revtex4}
\usepackage{latexsym}

\def\bs{\bar{\sigma}}
\def\rbar{\bar{r}}
\def\thetab{\bar{\theta}}
\def\bo{\bar{\Omega}}

\def\lb{\left(}
\def\rb{\right)}
\def\be {\begin{equation}}
\def\ee {\end{equation}  }
\def\beq{\begin{eqnarray}}
\def\eeq{\end{eqnarray}  }
\def\bi {\begin{itemize} }

\def\ei {\end{itemize}   }
\def\RE {I\kern-6pt R    }
\def\Z  {Z\kern-13pt Z   }
\def\be {\begin{equation}}
\def\ee {\end{equation}  }
\def\beq{\begin{eqnarray}}
\def\eeq{\end{eqnarray}  }

\def\eeq{\end{eqnarray}  }
\input epsf
\epsfclipon

\begin{document}


\title{An Axisymmetric Gravitational Collapse Code}
\author{Matthew W. Choptuik}
\affiliation{CIAR Cosmology and Gravity Program,
     Department of Physics and Astronomy,
     University of British Columbia,
     Vancouver BC, V6T 1Z1 Canada}
\author{Eric W. Hirschmann}
\affiliation{Department of Physics and Astronomy,
     Brigham Young University,
     Provo, UT 84604}
\author{Steven L. Liebling}
\affiliation{ Southampton College, Long Island University,
     Southampton, NY 11968}
\author{Frans Pretorius}
\affiliation{Theoretical Astrophysics,
     California Institute of Technology,
     Pasadena, CA 91125}

\begin{abstract}
We present a new numerical code designed to solve the Einstein
field equations for axisymmetric spacetimes. The long term
goal of this project is to construct a code
that will be capable of studying many problems of interest
in axisymmetry, including gravitational collapse, critical
phenomena, investigations of cosmic censorship, and head-on
black hole collisions. Our objective here is 
to detail the (2+1)+1 formalism we use to arrive at the corresponding
system of equations and the numerical methods we use to solve them.
We are able to obtain stable evolution, despite the singular
nature of the coordinate system on the axis, by enforcing
appropriate regularity conditions on all variables and by
adding numerical dissipation to hyperbolic equations. 
\end{abstract}


\maketitle


\section{Introduction}
In this paper we introduce a numerical code designed to
solve the Einstein field equations for axisymmetric spacetimes.
Even though the predominant focus in numerical
relativity in recent years has been to study situations of relevance to
gravitational wave
detection, and hence lacking symmetries, there are still numerous
interesting problems, both physical and computational, that can be
tackled with an axisymmetric code. The advantages of restricting
the full 3D problem to axisymmetry (2D) are that the complexity and
number of equations are reduced, as are the computational
requirements compared to solving a similar problem in 3D.
 
Prior numerical studies of axisymmetric spacetimes include
head-on black hole 
collisions~\cite{smarr,eppley,shapiro_teukolsky2,anninos,baker_et_al,anninos2},
collapse of rotating fluid 
stars~\cite{nakamura,stark_piran,nakamura_et_al,shibata},
the evolution of collisionless particles applied to study the stability
of star clusters~\cite{abrahams_cook} and the validity of cosmic 
censorship~\cite{shapiro_teukolsky}, evolution of gravitational
waves~\cite{garfinkle,alcubierre_etal}, black hole-matter-gravitational wave
interactions \cite{brandt_font_et_al,brandt_seidel1,brandt_seidel2,brandt_seidel3}, 
and the formation of black holes through gravitational wave collapse~\cite{abrahams2} 
and corresponding critical behavior at the threshold of formation~\cite{abrahams}.
 
Our goals for creating a new axisymmetric code are not only to
explore a wider range of phenomena than those studied before, but also  
to provide a framework to add adaptive mesh refinement (AMR) and
black hole excision to allow more thorough and detailed investigations than
prior works.
 
The outline of the rest of the paper is as follows. In Section~\ref{sec:formalism}
we describe the $(2+1)+1$ decomposition~\cite{maeda,maeda2} of spacetime that we
adopt to arrive at our system of equations. The $(2+1)+1$
formalism is the familiar ADM space + time decomposition (in this case $2+1$)
applied to a dimensionally reduced spacetime obtained by dividing out the
axial Killing vector, following a method devised by Geroch~\cite{geroch}.
In Section~\ref{sec:coords_and_vars} we specialize the equations to our
chosen coordinate system, namely cylindrical coordinates
with a conformally flat $2$--metric. At this stage we do not model
spacetimes with angular momentum, and we include a massless scalar field for the 
matter source. In Section~\ref{sec:apph} we discuss how we search for
apparent horizons during evolution.
In Section~\ref{sec:implementation} we describe our numerical implementation
of the set of equations derived in Section~\ref{sec:coords_and_vars}.
A variety of tests of our code are presented in Section~\ref{sec:test},
which is followed by conclusions in Section~\ref{sec:conclusion}.
Some details concerning our finite difference approximations, solution 
of elliptic equations via the multi-grid technique, and a spherically 
symmetric code used for testing purposes are given in Appendices A and B.
Unless otherwise specfied, we use the units and conventions adopted by
Misner, Thorne and Wheeler~\cite{MTW}.
\section{The (2+1)+1 Formalism}
\label{sec:formalism}
The most common approach in numerical relativity is to perform the so-called 
$3+1$, or ADM, split of the spacetime that one would like to evolve.  In 
this procedure, a timelike vector field is chosen together with spatial
hypersurfaces that foliate the spacetime.  If axisymmetry is assumed,
it is usually incorporated once the ADM decomposition has been done, and is 
reflected in the independence of the various metric and matter quantities
on the ignorable, angular coordinate, $\varphi$.  

Our approach, which follows Geroch~\cite{geroch} and 
Nakamura~{\it et al}~\cite{nakamura_et_al}, 
reverses this
procedure.  We assume axisymmetry from the outset and perform a reduction of 
the spacetime based on this assumption.  Once we have projected out the 
symmetry, we perform an ADM-like split (now a $2+1$ split) of the 
remaining $3$--manifold.  

More specifically, we begin with a $4$--dimensional spacetime metric 
\footnote{Note that we will use Greek letters to denote coordinates in the full 
$4$--manifold and Latin letters to denote coordinates in the reduced 
$3$--manifold.} 
$\gamma_{\mu \nu}$ on our manifold $\cal M$.  The axisymmetry is realized 
in the existence of a spacelike Killing vector
\be
X^\mu = \left(\frac{\partial}{\partial \varphi}\right)^\mu,
\ee
with closed orbits.
We define the projection operator $g_{\mu\nu}$, allowing us to project
tensors from the 4--dimensional manifold ${\cal M}$ with metric 
$\gamma_{\mu\nu}$ to a $3$--dimensional manifold ${\cal M}/S^1$ with metric
$g_{ab}$, as
\be
g_{\mu \nu} = \gamma_{\mu \nu} - \frac{1}{s^2} X_\mu X_\nu,
\ee
where $s$ is the norm of the Killing vector 
\be
s^2=X^\mu X_\mu.
\ee
With the definition of the vector 
\be
Y_\mu = \frac{1}{s^2} X_\mu,
\ee
the metric on the full, $4$--dimensional spacetime can be written as
\be
\gamma_{\mu \nu} = \left( \begin{array}{cc}
                          g_{ab} + s^2 Y_a Y_b  & s^2 Y_a \\
                          s^2 Y_b               & s^2
                          \end{array} \right).
\ee

Projecting, or dividing out the symmetry amounts to expressing 
$4$--dimensional quantities in terms of quantities on the $3$--manifold.
For instance, the connection coefficients ${}^{(4)} \Gamma^\lambda_{\mu \nu}$ 
associated with the $4$--dimensional metric are
\beq
{}^{(4)} \Gamma^\lambda_{\mu \nu} & = & {}^{(3)} \Gamma^\lambda_{\mu \nu} + \Omega^\lambda_{\mu \nu} 
\nonumber \\
 & = & {}^{(3)} \Gamma^\lambda_{\mu \nu}
+ \frac{1}{2} s^2 g^{\lambda \sigma} \left[ Y_\nu Z_{\mu \sigma}
                                           +Y_\mu Z_{\nu \sigma}
                                           -\partial_\sigma \left( \ln s^2 \right) Y_\mu Y_\nu
                                     \right] 
+ \frac{1}{2} Y^\lambda \left[ \partial_\mu \lb s^2 Y_\nu \rb + \partial_\nu \lb s^2 Y_\mu \rb \right],
\label{christof}
\eeq
where  $^{(3)}\Gamma_{\mu\nu}^\lambda$ are the connection coefficients 
constructed from the $3$--metric $g_{\mu\nu}$, and we have defined the antisymmetric tensor
\be
Z_{\mu \nu} = \partial_\mu Y_{\nu} - \partial_\nu Y_{\mu}.
\ee
Notice that $Z_{\mu \nu}$ is an intrinsically $3$--dimensional object,
in that $Z_{\mu \nu} X^\mu = Z_{\mu \nu} X^\nu = 0$.
With some algebra, the Ricci tensor on the $4$--manifold, ${}^{(4)}R_{\mu\nu}$, can now be 
written as the Ricci tensor on the reduced space, ${}^{(3)}R_{\mu\nu}$, together with 
additional terms involving fields coming from the dimensional reduction
\beq
{}^{(4)}R_{\mu\nu} = {}^{(3)}R_{\mu\nu} + D_{\lambda}
   \Omega^{\lambda}_{\mu\nu} -
   D_{\mu}\Omega^{\lambda}_{\nu\lambda}
   + \Omega^{\lambda}_{\mu\nu} \Omega_{\lambda\sigma}^\sigma -
     \Omega^{\sigma}_{\mu\lambda} \Omega^{\lambda}_{\nu\sigma},
\label{ricci_4}
\eeq
where $D_{\mu}$ is the covariant derivative on the $3$ dimensional
manifold.
Expressed in terms of $s$, $Y_a$, $Z_{ab}$ and ${}^{(3)}R_{ab}$, the components of 
(\ref{ricci_4}) are
\beq
{}^{(4)}R_{\varphi\varphi} & = & {1\over 4} s^4 Z_{bc}Z^{bc}
         - s D^{a} D_{a} s,  \label{eq:r_phiphi} \\
{}^{(4)}R_{\varphi a} & = & {1\over 2s} D^{c}(s^3 Z_{ac}) 
       + Y_{a}\left[ {1\over 4} s^4 Z_{bc}Z^{bc}
      - s D^{a} D_{a} s  \right],  \label{eq:r_rhia}\\
{}^{(4)}R_{ab} & = & {}^{(3)}R_{ab} - {1\over s} \,
     D_{a} D_{b}s
      - {1\over 2} s^2 Z_{ac}Z_{b}{}^{c}   
          - {1\over s} \,
    D^{c}\bigl[ s^3 Z_{c(a} \bigr] Y_{b)}
      + Y_{a} Y_{b} \left[ {1\over 4} s^4 Z_{bc}Z^{bc}
         - s D^{a} D_{a} s \right]. \label{eq:r_ab}
\eeq
Taking the trace of (\ref{ricci_4}), and using the definitions 
described above, gives the decomposition of the Ricci scalar as
\be
{}^{(4)}R = {}^{(3)}R - {2\over s} D^{a} D_{a}s -
     {1\over 4} s^2 Z_{bc}Z^{bc}.
\label{eq:ricci}
\ee
The $4$--dimensional Einstein equations, with a stress-energy tensor
$T_{\mu\nu}$, are
\be
{}^{(4)}R_{\mu\nu} - \frac{1}{2} {}^{(4)}R \gamma_{\mu\nu} = 8\pi T_{\mu\nu}.
\label{eq:einstein}
\ee
Using equations (\ref{eq:r_phiphi}-\ref{eq:ricci}), we can write the 
Einstein equations as
\beq 
D^{a}D_{a} s & = & - {w_{a}w^{a} \over 2s^3} 
                   - {8\pi \over s}
                        \left(   T_{\varphi\varphi}
                               - {1\over2} s^2 T_{\lambda}{}^{\lambda}
                       \right), \label{eq:s_eom}                          \\
D_{[a} w_{b]} & = & 8 \pi \, s \, \epsilon_{ab}{}^{c} \, T_{c\varphi},
                      \label{eq:w_eom} \\
{}^{(3)}R_{ab} & = &    {1\over s} D_{a} D_{b}s
              + {1\over 2} s^{2} Z_{ac}Z_{b}{}^{c}                        
              + 8\pi \left( T_{\mu\nu} g^{\mu}{}_{a} g^{\nu}{}_{b}
              - {1\over2} g_{ab} T_{\lambda}{}^{\lambda} \right),
                \label{eq:r_eom} 
\eeq 
where we have introduced the twist $w_\mu$ of the Killing vector
\be
w_\mu = {s^4\over2} \, \epsilon_{\mu\nu\lambda\sigma} 
             Y^\nu Z^{\lambda\sigma},
\ee
and the four and three dimensional Levi-Civita 
symbols $\epsilon_{\mu\nu\lambda\sigma}$ 
and $\epsilon_{abc}$, respectively. The twist vector $w_\mu$ is
intrinsically $3$--dimensional, {\it i.e.} $w_\mu X^\mu=0$. Furthermore,
$w_a/s^3$ is divergence free
\be
D^a\left[\frac{w_a}{s^3}\right]=0. \label {eq:div_w_eom}
\ee

At this point, the first reduction is essentially done.  
Equation (\ref{eq:r_eom}) can be viewed as the $3$--dimensional
Einstein equations, coupled to the projection of the stress-energy
tensor $T_{\mu\nu}$ and to induced ``matter fields'' $s$ and $w_{a}$ 
(or $Z_{ab}$). The equations of motion for $s$ and $w_{a}$ are 
given by (\ref{eq:s_eom}) and (\ref{eq:w_eom})
respectively; additional equations of motion will need to be specified for
whatever true matter fields one incorporates into the system.
This procedure so far is completely analogous to the Kaluza-Klein reduction 
from five to four dimensions, in which the $5$--dimensional geometry becomes 
gravity in $4$ dimensions coupled to electromagnetism and a scalar field
\cite{kk}.
Here, however, the reduced $3$--manifold has no dynamics and we have the
rather appealing picture of having ``divided out" the dynamics of the 
four dimensional gravitational system and reinterpreted the two degrees of 
freedom in the gravitational field as scalar ($s$) and ``electromagnetic'' 
($Z_{ab}$, or $w_a$) degrees of freedom.\footnote{Recall 
that the electromagnetic 
field in $3$ dimensions has only a single degree of freedom as compared 
to the two degrees of freedom in $4$ dimensions.}  

We now perform the ADM split of the remaining spacetime.  This is 
done by first foliating the $3$--dimensional spacetime into a series of spacelike
hypersurfaces with unit, timelike normal vector $n^a$. Then, similar to the dimensional
reduction above, we decompose quantities into components orthogonal and
tangent to $n^a$, using the projection tensor $h_{ab}$
\be
h_{ab} = g_{ab} + n_a n_b.
\ee
We now define the components of $n^a$, and the induced $2$--dimensional metric 
$h_{AB}$ \footnote {We use upper-case Latin indicies to denote $2$-dimensional tensor 
components.}, using the following decomposition of the $3$--metric
\be
g_{ab} dx^{a} dx^{b} = -\alpha^2 dt^2 + h_{AB} (dx^{A} +
   \beta^{A}dt) (dx^{B} + \beta^{B}dt),
\ee
where $\alpha$ is the lapse function and $\beta^A$ the shift vector.
The gravitational equations now become the evolution equations
for the components of the $2$--metric $h_{AB}$ and the $2$--extrinsic curvature $K_{AB}$
\beq 
\partial_{t} h_{AB} & = & -2\alpha K_{AB} +\Delta_A \beta_B +\Delta_B \beta_A \label{eq:hdot} \\
\partial_{t} K_{AB} 
  & = &     \beta^{C} \Delta_{C} K_{AB} 
          + K_{AC} \Delta_B \beta^C
          + K_{BC} \Delta_A \beta^C       
          + \alpha \bigl[ K K_{AB} 
          + \, ^{(2)}R_{AB} \bigr]              \nonumber \\ 
  &   &   - 2\alpha K_{AC}K_{B}{}^{C} 
          - \Delta_{A}\Delta_{B}\alpha
          - \alpha\, ^{(3)}R_{AB},           \label{eq:kdot}
\label{eq:k_AB_dot}
\eeq 
the Hamiltonian constraint equation
\be
^{(2)}R - K^{A}{}_{B}K^{B}{}_{A} + K^2 
    =   \, ^{(3)}R + 2\, ^{(3)}R_{ab}n^a n^b  \label{eq:hc}      
\ee
and the $\rho$ and $z$ momentum constraint equations
\be
\Delta_{A} K_{B}{}^{A} - \Delta_{A}K  
    =   -\, ^{(3)}R_{cd} n^d h^c{}_B.         \label{eq:mc}
\ee
In the above, $\Delta_{A}$ is the covariant derivative compatible with 
the $2$--metric $h_{AB}$, $K\equiv K_A{}^A$, and ${}^{(2)}R_{AB}$
and ${}^{(2)}R$ are the $2$--dimensional Ricci tensor and Ricci scalar,
respectively.
Note that because gravity in $3$ dimensions has no propagating degrees 
of freedom, the constraint equations fix the $3$--dimensional geometry 
completely. Thus, if desired, one can use the constraint equations 
(\ref{eq:hc}-\ref{eq:mc}) instead of the evolution equations
(\ref{eq:hdot}-\ref{eq:kdot}) to solve for $h_{AB}$. 
The freely specifiable degrees of 
freedom of the $4$--manifold are encoded
in $s$ and $w_{a}$, which are evolved using (\ref{eq:s_eom}), (\ref{eq:w_eom}) and 
(\ref{eq:div_w_eom}). Note that (\ref{eq:w_eom}) and (\ref{eq:div_w_eom}) 
constitute four equations for the three components of $w_a$---the purely spatial
have four equations for the three components of $w_a$.  The purely spatial
part of (\ref{eq:w_eom}) is, in $3+1$ language, the angular momentum constraint
equation and only needs to be solved at the initial time in a free
evolution of $w_a$.
We note that the restricted class of axisymmetric spacetimes having no 
angular momentum (rotation) is characterized by the existence of 
a {\em scalar twist}, $w$, such that $w_a\equiv w_{,a}$. 
In the vacuum case, $w$ generally represents odd parity
gravitational waves, while $s$ encodes even parity, or Brill waves\cite{brill}.
We further note that this class includes the special case $w_a \equiv 0$,
which will be the focus of our discussion below.

\section{Coordinate System, Variables and Equations}
\label{sec:coords_and_vars}
In this section we describe a particular coordinate system and
set of variables which, in the context of 
the formalism described in the previous section, provides us with the concrete
system of partial differential equations that we solve numerically.
We also detail the outer boundary conditions we use,
and the on-axis regularity conditions necessary to obtain
smooth solutions to these equations. 

We only consider spacetimes with zero angular momentum,
and no odd-parity gravitational waves; therefore $w_a=0$.
We choose a conformally flat, cylindrical coordinate system
for the $2$--metric
\be
h_{AB} dx^A dx^B = \psi(\rho,z,t)^4 ( d\rho^2 + dz^2 ). \label{HAB_def}
\ee
This choice for $h_{AB}$ exhausts the coordinate freedom we have 
to arbitrarily specify the two components of the shift 
vector---$\beta^\rho(\rho,z,t)$ and $\beta^z(\rho,z,t)$. 
In order to maintain the form~(\ref{HAB_def}) during an evolution, we 
use the momentum constraints, which are elliptic equations, to solve for $\beta^\rho$ 
and $\beta^z$ at each time step.
The Hamiltonian constraint provides a third elliptic equation that we can use
to solve for the conformal factor $\psi$.
For a slicing condition, we use maximal slicing of
$t={\rm const}.$ hypersurfaces {\em in the $4$--dimensional manifold}---{\it i.e.} 
we impose ${}^{(3)}K = 0$, where ${}^{(3)}K$ is the trace of the extrinsic
curvature tensor of $t={\rm const}.$ slices of $\gamma_{\mu\nu}$.  
This condition (specifically $\partial {}^{(3)}K/\partial t = 0$) 
gives us an elliptic equation for the lapse.

Instead of directly evolving the norm of the Killing vector, $s$, we
evolve the quantity $\bs$, defined by
\be
s = \rho \psi^2 e^{\rho \bar \sigma},
\ee
and furthermore, we convert the resultant evolution equation for $\bs$
(\ref{eq:s_eom}) to one that is first order in time by defining the quantity 
$\bo$, which is 
``conjugate'' variable to $\bs$, via 
\beq
\rho\bar \Omega &=& - 2K_\rho{}^\rho - K_z{}^z   \nonumber \\
                &=& - \frac{3}{2} \, n^a (\ln s)_{,a} 
                    + \frac{\beta^z_{,z}-\beta^\rho_{,\rho}}{2\alpha}. \label{eq:bo_def}
\eeq
Part of the motivation behind using $\bs$ and $\bo$ as
fundamental variables is to simplify the enforcement of on-axis regularity.
In particular, regularity as $\rho \to 0$ implies that
$\bs$ and $\bo$ must exhibit leading order behavior of the form 
$\bs=\bs_1(z,t)\rho + O(\rho^3)$ and $\bo=\bo_1(z,t)\rho + O(\rho^3)$ respectively, 
and experience has shown it to be easier to enforce such conditions, than
to enforce the 
leading order behavior of $s$ (or its time derivative) near the axis, 
which is $s=s_2(z,t)\rho^2 + O(\rho^4)$. 

As mentioned previously, the only matter source we 
currently incorporate is a massless scalar field 
$\Phi(\rho,z,t)$, which satisfies the usual $4$--dimensional wave equation
$\Box\Phi=0$. We convert this equation to first-order-in-time form by defining a
conjugate variable $\Pi$:
\be
\Pi \equiv \psi^2 \, n^a \Phi_{,a} .
\ee
The stress-energy tensor for the scalar field is
\be
T_{\mu\nu} =  2 \Phi_{,\mu} \Phi_{,\nu}
             - \gamma_{\mu\nu} \Phi^{,\mu} \Phi_{,\mu}.
\ee

Using all of the above definitions and restrictions within the formalism
detailed in the previous section, we end up with the following system
of equations that we solve with our numerical code, described
in the next section. The maximal slicing condition results in the following
elliptic equation for $\alpha$
\begin{eqnarray}
2 \left( \rho \alpha_{,\rho} \right)_{,\rho^2} + \alpha_{,zz}
        + \alpha_{,\rho} \Bigl(  
                                  2 \frac{ \psi_{,\rho} }{ \psi }
                                + \left( \rho \bs \right)_{,\rho}
                         \Bigr)
        + \alpha_{,z   } \Bigl(  
                                  2 \frac{ \psi_{,z   } }{ \psi }
                                + \left( \rho \bs \right)_{,z   }
                         \Bigr)
\qquad\qquad\qquad\qquad\qquad
\nonumber\\
   - \frac{\psi^4}{2\alpha}
       \Bigl[  
                 (  \beta^\rho{}_{,\rho} - \beta^z{}_{,z}      )^2
               + (  \beta^\rho{}_{,z   } + \beta^z{}_{,\rho} )^2 
       \Bigr]
   - \frac{\psi^4}{6\alpha} 
       \Bigl[
                 2 \alpha \rho \bo
               + \beta^\rho{}_{,\rho}
               - \beta^z{}_{,z}
       \Bigr]^2
   = 16 \pi \alpha \Pi^2 .
\label{eq:cons_alpha}
\end{eqnarray}
The Hamiltonian constraint gives an elliptic equation for $\psi$
\begin{eqnarray}
          8 \frac{ \psi_{,\rho \rho} }{ \psi}
       +  8 \frac{ \psi_{,z    z   } }{ \psi}
       + 16 \frac{ \psi_{,\rho^2} }{ \psi }
       +  8 \left( \rho \bs \right)_{,\rho}
              \frac{ \psi_{,\rho  } }{ \psi }
       +  8 \left( \rho \bs \right)_{,z   }
              \frac{ \psi_{,z     } }{ \psi }
\qquad\qquad\qquad\qquad\qquad\qquad\qquad
\nonumber\\
   + {\psi^4 \over 2 \alpha^2}
        \Bigl[
                   (  \beta^\rho{}_{,\rho} - \beta^z{}_{,z}      )^2
                 + (  \beta^{\rho}{}_{,z}  + \beta^z{}_{,\rho}   )^2 
        \Bigr]
   + {\psi^4 \over 6\alpha^2}
        \Bigl[ 
                   2\alpha\rho\bar{\Omega}
                 + \beta^\rho{}_{,\rho}
                 - \beta^z{}_{,z}   
        \Bigr]^2
\quad\qquad\qquad
\nonumber\\
\qquad
 \, = \,  
            - 16 \pi \left( \Pi^2 + \Phi_{,\rho}{}^2 + \Phi_{,z}{}^2 \right)
            -  6 \bigl( \rho^2 \left(\rho \bs\right)_{,\rho} \bigr)_{,\rho^3}
\nonumber\\
            -  2 \bigl( \left( \rho \bs \right)_{,\rho} \bigr)^2
            -  2 \bigl( \rho \bs \bigr)_{,zz}
            -  2 \bigl( \left( \rho \bs \right)_{,z   } \bigr)^2 .
\label{eq:cons_psi}
\end{eqnarray}
The $\rho$ and $z$ momentum constraints, which provide elliptic equations
that we use to solve for $\beta^\rho$ and $\beta^z$, are
\begin{eqnarray}
   \frac{2}{3}\beta^{\rho}{}_{,\rho\rho}
 + \beta^{\rho}{}_{,zz}
 + \frac{1}{3} \beta^z{}_{,z\rho}
 -\frac{2 \alpha \rho }{ 3 } \Bigl[
                                     6 \bo \frac{\psi_{,\rho}}{ \psi }
                                   + \bo_{,\rho}
                                   + 3 \bo \left(\rho \bs   \right)_{,\rho}
                             \Bigr] 
 - \frac{8}{3} \, \alpha \bar{\Omega}
\qquad\qquad\qquad\qquad\qquad\qquad
\nonumber\\
 - \frac{2}{3} \Bigl[     
                           \frac{ \alpha_{,\rho} }{ \alpha }
                       - 6 \frac{ \psi_{,\rho}   }{ \psi   }
               \Bigr] 
     \bigl(
               \beta^\rho{}_{,\rho}
             - \beta^z{}_{,z}
     \bigr)
 - \Bigl[      
               \frac{ \alpha_{,z} }{ \alpha }
           - 6 \frac{ \psi_{,z}   }{ \psi   } 
           - \left( \rho\bs \right)_{,z}
   \Bigr]
   \bigl(    
             \beta^\rho{}_{,z}
           + \beta^z{}_{,\rho}
   \bigr)
 \; = \; - 32\pi {\alpha \over \psi^2} \Pi_{,\rho} , 
\label{eq:cons_betarho}
\end{eqnarray}
and
\begin{eqnarray}
   \beta^z{}_{,\rho\rho}
 + \frac{4}{3}\beta^z{}_{,zz}
 - \frac{1}{3} \beta^{\rho}{}_{,z\rho}
 - \frac{2 \alpha\rho}{3} \Bigl[ 
                                   6 \bo \frac{\psi_{,z}}{\psi}
                                 + \bo_{,z}
                                 + 3 \bo (\rho \bs)_{,z}
                          \Bigr]
\qquad\qquad\qquad\qquad\qquad\qquad\qquad\quad\quad
\nonumber\\
   + \frac{4}{3} \, \Bigl[   
                               \frac{ \alpha_{,z} }{ \alpha }
                           - 6 \frac{ \psi_{,z}   }{ \psi   } 
                           - \frac{3}{2} \left( \rho \bs \right)_{,z}
                    \Bigr]
       \bigl( \beta^\rho{}_{,\rho} - \beta^z{}_{,z} \bigr)
   + \Bigl[
             \frac{ 2 \alpha }{ \psi^6} \Bigl(
                                                \frac{\rho \psi^6}{\alpha}
                                        \Bigr)_{,\rho^2} \, 
           + \left( \rho \bs \right)_{,\rho}
     \Bigr]
       \bigl( \beta^\rho{}_{,z} + \beta^z{}_{,\rho} \bigr)
 \; = \; - 32\pi {\alpha \over \psi^2} \Pi_{,z} .
\label{eq:cons_betaz}
\end{eqnarray}
The definition of $\bo$ in Eq.~(\ref{eq:bo_def}) gives an 
evolution equation for $\bs$ (where the overdot 
denotes partial differentiation with respect to $t$)
\begin{equation}\label{eq:evol_bs}
\dot{\bs} =
        2 \beta^{\rho} \left( \rho\bs \right)_{,\rho^2} + \beta^z
        \bs_{,z}
     -  \alpha\bar{\Omega}
     - \Bigl( {\beta^{\rho} \over \rho} \Bigr)_{,\rho}.
\end{equation}
The evolution equation for $\bo$ is
\begin{eqnarray}
\dot{\bar{\Omega}} & = & 
     2 \beta^{\rho} \left(\rho\bar{\Omega}\right)_{,\rho^2}
    +  \beta^z \bar{\Omega}_{,z}
    - {1\over 2\alpha\rho} \left(
                                   \beta^{z}{}_{,\rho}{}^2
                                 - \beta^{\rho}{}_{,z}{}^2
                                 \right)
    + {1\over \psi^4} \Bigl( {\alpha_{,\rho} \over \rho} \Bigr)_{,\rho}
\nonumber\\
& & + {\alpha \over \psi^6}\Bigl(
               {(\psi^2)_{,\rho} \over \rho} \Bigr)_{,\rho}
    - {2\alpha \over \psi^4}
        \Bigl(   
                 4 \frac{ \psi_{,\rho^2} }{ \psi }
               + \left( \rho \bs \right)_{,\rho^2}
        \Bigr)
        \Bigl(   
                 \frac{\alpha_{,\rho}    }{ \alpha }
               + \frac{ 2\psi_{,\rho} }{ \psi   }
        \Bigr)
\nonumber\\
& & - {\alpha \over \psi^4} \Bigr[
                                 \bs_{,z} \Bigl(
                                                  \frac{ \alpha_{,z} }{ \alpha}
                                                + \frac{ 2 \psi_{,z} }{ \psi  }
                                          \Bigr)
                               + \rho \bs_{,z}{}^2
                               + \bs_{,zz}
                           \Bigr]
    + 64\pi{\alpha\over \psi^4} \rho (\Phi_{,\rho^2})^2.
\label{eq:evol_bo}
\end{eqnarray}
We also have an evolution equation for $\psi$, which 
we optionally use instead of the Hamiltonian constraint (\ref{eq:cons_psi}) 
to update $\psi$
\begin{equation}\label{eq:evol_psi}
\dot \psi =  \psi_{,z}\beta^z+\psi_{,\rho}\beta^\rho+
             \frac{\psi}{6}\left(2\beta^\rho_{,\rho}+\beta^z_{,z}+
             \rho\alpha\bar{\Omega}\right).
\end{equation}
The definition of $\Pi$ and the wave equation for $\Phi$ give
\begin{equation}\label{eq:evol_phi}
\dot{\Phi} =
       \beta^{\rho} \Phi_\rho + \beta^z \Phi_z
            + {\alpha \over \psi^2 } \Pi,
\end{equation}
and
\begin{eqnarray}
\dot{\Pi} & = & 
       \beta^{\rho} \Pi_{,\rho} + \beta^z \Pi_{,z}
     + {1\over3} \Pi \left(   \alpha\rho\bar{\Omega}
                            + 2 \beta^{\rho}_{,\rho}
                            + \beta^z_{,z}
                     \right)
\nonumber\\
& &  + \frac{1}{\psi^4} \left[
                         2\left( \rho \alpha \psi^2 \Phi_\rho \right)_{,\rho^2}
                        + \left(      \alpha \psi^2 \Phi_z    \right)_{,z}
                              \right]
     + \frac{\alpha}{\psi^2} \left[
                                     \left( \rho \bs \right)_{,\rho} \Phi_\rho
                                   + \left( \rho \bs \right)_{,z   } \Phi_z
                                  \right].
\label{eq:evol_pi}
\end{eqnarray}

To complete the specification of our system of equations, we need to
supply boundary conditions. In our cylindrical coordinate system,
where $\rho$ ranges from $\rho=0$ to $\rho=\rho_{\rm max}$
and $z$ ranges from $z_{\rm min}$ to $z_{\rm max}$,
we have two distinct boundaries: the physical outer boundary
at $\rho=\rho_{\rm max}$, $z=z_{\rm min}$, and $z=z_{\rm max}$;
and the axis, at $\rho=0$. Historically, the axis presented a stability 
problem in axisymmetric codes. We solve this problem by enforcing 
regularity on the axis, and, as described in Section
\ref{sec:implementation}, adding
numerical dissipation to evolved fields.

The regularity conditions can be obtained by inspection
of the equations in the limit $\rho \rightarrow 0$, or more formally,
by transforming to Cartesian coordinates and demanding that components
of the metric and matter fields be regular and single valued throughout
\cite{bardeen}. 
Garfinkle and Duncan~\cite{garfinkle} have further proposed that 
in order to ensure smoothness on the axis, one  
should use quantities that have either even or odd power series expansions
in $\rho$ as $\rho \to 0$, but which do not vanish faster than $O(\rho)$.
It is interesting that
the quantities which we found to work best also obey this requirement.
As discussed earlier, the particular choice
of $\bs$ and $\bo$ as fundamental variables was partly
motivated by regularity concerns. The results are
\beq
\alpha_{,\rho}(0,z,t) & = & 0
\label{eq:bc_alpha}
\\
\psi_{,\rho}(0,z,t) & = & 0
\\
\beta^z{}_{,\rho}(0,z,t) & = & 0
\\
\beta^\rho{}(0,z,t) & = & 0
\label{eq:bc_betarho}
\\
\bs(0,z,t) & = & 0
\\
\bo(0,z,t) & = & 0
\\
\Phi_{,\rho}(0,z,t) & = & 0
\\
\Pi_{,\rho}(0,z,t) & = & 0
\eeq
At the outer boundary, we enforce asymptotic flatness by requiring
\beq
\lim_{r\to\infty}
\alpha(r,t) & = & 1 + \frac{C(t)}{r} + O(r^{-2})
\label{eq:obc_alpha}
\\
\lim_{r\to\infty}
\psi(r,t)   & = & 1 + \frac{D(t)}{r}  + O(r^{-2}) 
\\
\lim_{r\to\infty}
\beta^z(r,t)   & = & \frac{E(t)}{r} +  O(r^{-2})  
\\
\lim_{r\to\infty}
\beta^\rho(r,t)   & = & \frac{F(t)}{r}  + O(r^{-2}),
\label{eq:obc_betarho}
\eeq
for undetermined functions $C(t)$, $D(t)$, $E(t)$, $F(t)$, 
and $r^2\equiv \rho^2+z^2$.
These latter relations are converted to mixed (Robin) boundary 
conditions (see Appendix A for details) and then are imposed at the 
outer boundaries of the computational domain: $\rho = \rho_{\rm max}$,
$z = z_{\rm max}$, and $z=-z_{\rm max}$.
We have also experimented with the use of Dirichlet conditions on
$\alpha,\beta^\rho$ and $\beta^z$ at the outer
boundaries (specifically
$\alpha=1$ and $\beta^\rho=\beta^z=0$ there), and have found that these
work about as well as the Robin conditions.
For the scalar field, we assume that near the outer boundary 
we can approximate the field as purely radially outgoing, and require
\be
\left( r \Phi \right)_{,t} + \left(r \Phi \right)_{,r} = 0.
\ee
For scalar field configurations far from spherical symmetry,
this approximation suffers and reflections are relatively large.
However, in general, the reflections do not grow and are somewhat damped.
For the other two evolved quantities, $\bs$ and $\bo$, we use
this same naive condition for lack of any better, more physically
motivated conditions. While this condition proves to be stable with damped
reflections, a better condition is sought and this issue remains
under investigation.

For initial conditions, we are free to set 
$(\bs(0,\rho,z),\bo(0,\rho,z),\Phi(0,\rho,z),\Pi(0,\rho,z))$.  Once the 
free data is chosen, we 
then use the constraint and slicing equations to determine 
$(\alpha(0,\rho,z),\psi(0,\rho,z),\beta^z(0,\rho,z),\beta^\rho(0,\rho,z))$.
Specifically, we define a general pulse shape
\be
G_X(\rho,z)   =  A_X \exp \left[ -\left(\frac{
                                  \sqrt{  \left(\rho-\rho_X\right)^2
                                     +\epsilon_X \left(z-z_X\right)^2}
                                    - R_X} {\Delta_X}\right)^2 \right]
\label{G_X}
\ee
characterized by six parameters
$(A_x, \rho_X, \epsilon_X, z_X, R_X, \Delta_X)$ and then 
choose initial data of the form
\beq
\bs(0,\rho,z)  & = & \rho \, G_{\bs}(\rho,z)
\cr
\bo(0,\rho,z)  & = & \rho \, G_{\bo}(\rho,z)
\cr
\Phi(0,\rho,z) & = & G_{\Phi}(\rho,z)
\cr
\Pi(0,\rho,z)  & = & 0.
\eeq
For $\epsilon_X=1$, these pulses are Gaussian, spherical shells 
centered at $(\rho_X,z_X)$ with radius $R_X$ and pulse width $\Delta_X$.
For $\epsilon_X=1$ and $(\rho_X,z_X)=(0,0)$, the pulses are spherical.
The factor of $\rho$ in the initial data for $\bs$ and $\bo$
ensures the correct behavior on axis for regularity.
For the evolutions presented here, we let $\Pi=0$ so that the initial
configuration represents 
a moment of time symmetry. We note, however, that we are also able 
to generate and evolve approximately ingoing initial data.

\section{Finding Apparent Horizons}
\label{sec:apph}
In this section we describe the equation and technique 
we use to search for apparent horizons (AHs) within $t={\rm const.}$ 
spatial slices of the spacetime (see \cite{nakamura,thornburg,alcubierre_et_al}
for descriptions of some of the methods available to find
AHs in axisymmetry).
We restrict our search to isolated,
simply connected AHs. In axisymmetry, such an AH can be described by
a curve in the $(\rho,z)$ plane, starting and ending on the axis
at $\rho=0$. We define the location of the AH
as the level surface $F=0$, where
\begin{equation}\label{fs2}
F = \rbar - R(\thetab),
\end{equation}
and
\begin{eqnarray}
\rbar &\equiv& \sqrt{\rho^2+(z-z_0)^2}, \\
\rbar \sin \thetab &\equiv& \rho, \\
\rbar \cos \thetab &\equiv& (z-z_0).
\end{eqnarray}
The AH is the outermost, marginally trapped surface; hence, we want
to find an equation for $R(\thetab)$ such that the outward null
expansion normal to the surface $F=0$, is zero.
To this end, we first construct the unit spatial vector $s^a$, normal to 
$F={\rm const.}$
\begin{equation}\label{sadef}
s^a=\frac{g^{ab} F_{,b}}
          {\sqrt{g^{cd} F_{,c}F_{,d}}}.
\end{equation}
Then, using $s^a$ and the $t={\rm const.}$ hypersurface normal vector $n^a$,
we construct future-pointing outgoing ($+$) and ingoing ($-$) null
vectors as
\begin{equation}\label{null}
\ell^a_{\pm}=n^a \pm s^a.
\end{equation}
The normalization of the null vectors is (arbitrarily)
$\ell^a_+ \ell_{-a}=-2$.
The outward null expansion $\theta_+$ is then the divergence of $\ell^a_+$ projected 
onto $F={\rm const.}$
\begin{equation}\label{exp1}
\theta_+ = \left(g^{ab}-s^a s^b \right) \nabla_b \ell_{+a}.
\end{equation}
Using the definition of the extrinsic curvature $K_{AB}$, and substituting
(\ref{null}) into (\ref{exp1}), we arrive at the familiar form for the
null expansion when written in terms of ADM variables
\begin{equation}\label{exp2}
\theta_+ = s^A s^A K_{AB} + \Delta_A s^A - K.
\end{equation}
Note that because the normalization of $\ell^a_+$ is arbitrary,
so (to some extent) is that of $\theta_+$. The
above normalization is chosen so that $\theta_+$ measures
the fractional rate of change of area with time measured by an
observer moving along $n^a$.

Substituting (\ref{fs2}) and (\ref{sadef}) into (\ref{exp2}), and setting $\theta_+=0$, 
we are left with an
ordinary differential equation for $R(\thetab)$. This equation takes the following form,
where a prime $'$ denotes differentiation with respect to $\thetab$:
\begin{equation}\label{rpp_eqn}
R''(\thetab) +
G(R'(\thetab),R(\thetab),g_{ab},g_{ab,\rho},g_{ab,z}) = 0.
\end{equation}
$G$ is a rather lengthy function of its arguments, non-linear
in $R$ and $R'$; for brevity we do not display it explicitly.
All of the metric functions and their gradients appearing 
in (\ref{rpp_eqn}) are evaluated along a given curve of integration,
and hence are implicitly functions of $\thetab$.
$\thetab$ ranges from $0$ to $\pi$, 
and regularity of the surface $F=0$ about the axis requires
$R'(0)=R'(\pi)=0$. Integration of (\ref{rpp_eqn}) therefore
proceeds by specifying $R$ at $\thetab=0$ (for instance), and
then ``evolving'' $R$ until either $\thetab=\pi$, or $R$
diverges at some value of $\thetab<\pi$. 
If an AH exists, and assuming
$z_0$ is inside the AH, then the AH can be found by searching for
the (locally) unique\footnote{In a
general collapse scenario, multiple inner horizons could be
present, which would also satisfy (\ref{rpp_eqn}) augmented
with the conditions $R'(0)=R'(\pi)=0$. We want the
outermost of these surfaces.} initial value $R(0)=R_0$ such
that integration of (\ref{rpp_eqn}) ends at $R(\pi)=R_\pi$,
with $R'(\pi)=0$ and $R_\pi$ finite. 
For $R(0)$ slightly larger than $R_0$ (outside the AH), 
the integration will end at $\thetab=\pi$ with $R'(\pi)>0$, 
indicating an irregular point on the surface; similarly,
for $R(0)$ slightly smaller than $R_0$ (inside the AH) the integration
will end with $R'(\pi)<0$.
Therefore, if we can find a reasonable bracket about
the unknown $R_0$, we can use a bisection search to find
$R_0$. Currently, we find a bracket to search by testing
a set of initial points, equally spaced in $z$ at intervals of
$3\Delta z$. This seems to work well in most situations,
and the search is reasonably fast.

We use a second-order Runge-Kutta method to integrate equation
(\ref{rpp_eqn}). The metric functions appearing in $G$ are
evaluated using bilinear interpolation along the curve.

\section{Implementation}
\label{sec:implementation}
In this section we describe the numerical code that we have
written to solve the equations listed in Section
\ref{sec:coords_and_vars}.
Some details are deferred to Appendix \ref{app:solver}.

We use a uniform grid of size $N_\rho$ points in $\rho$ by $N_z$ points in $z$, with
equal spacing $\Delta\rho=\Delta z=h$ in the $\rho$ and $z$ directions.
The value of a function $f$ at time level $n$
and location $(i,j)$ within the grid, corresponding to coordinate 
$(\rho,z,t)=
((i-1)\Delta\rho, (j-1)\Delta_z + z_{\rm min},n\Delta t)$, is denoted by
$f^n_{i,j}$. For the temporal discretization scale we use $\Delta t=\lambda h$, 
where $\lambda$ is the Courant factor, which 
for the type of differencing we employ, should be less than one
for stability; typically we use $\lambda=0.3$. 
The hyperbolic equations (\ref{eq:evol_bs}-\ref{eq:evol_pi})
are discretized using a second-order accurate
Crank-Nicholson type scheme, whereby we define two time
levels, $t$ and $t+\Delta t$, and obtain our finite difference stencils
by expanding in Taylor series about $t=t+\Delta t/2$. 
This gives the following second-order accurate approximation to the first
derivative of $f$ with respect to time
\begin{equation}\label{ref_cn_time}
\frac{f^{n+1}_{i,j}-f^{n}_{i,j}}{\Delta t} = 
\frac{\partial f(\rho,z,t)}{\partial t} \bigg\vert_{t=t+\Delta t/2} +
O(\Delta t^2)
\end{equation}
Second-order accurate approximations to functions and spatial derivative
operators at $t=t+\Delta t/2$ are obtained by averaging the corresponding
quantity, $Q$, in time:
\begin{equation}\label{ref_cn_spat}
\frac{Q^{n}_{i,j}+Q^{n+1}_{i,j}}{2}
=
Q(t+\Delta t/2,\rho,z) + O(\Delta t^2)
\end{equation}
Thus, after discretization of the evolution equations using (\ref{ref_cn_time})
and (\ref{ref_cn_spat}),
function values are only referenced at times $t$ and $t+\Delta t$,
even though the stencils are centered at time $t+\Delta t/2$.
Specific forms for all the finite difference stencils that we use can
be found in Appendix \ref{app:solver}. 

\begin{figure}
\centerline{\epsfxsize=8cm \epsffile{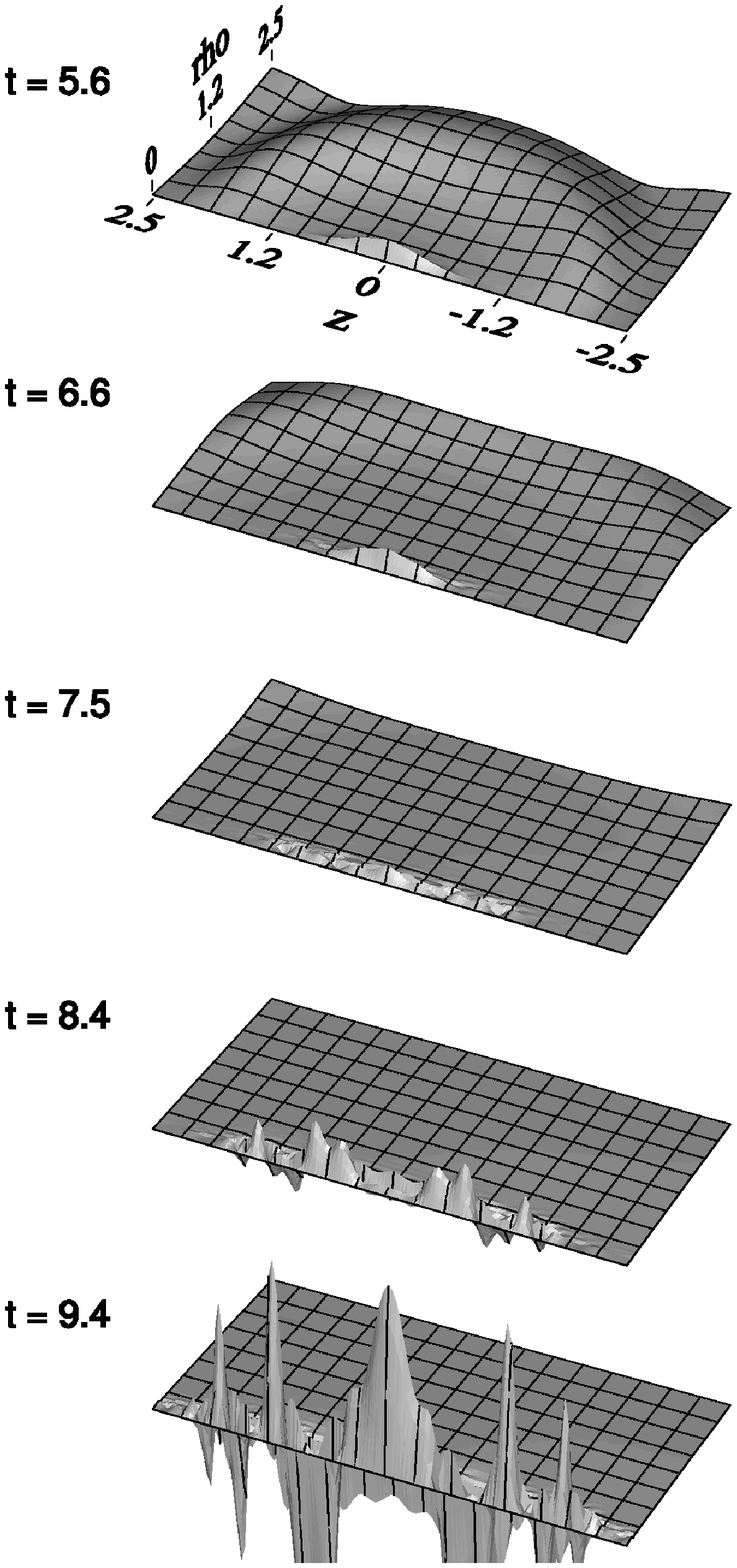} 
            \epsfxsize=8cm \epsffile{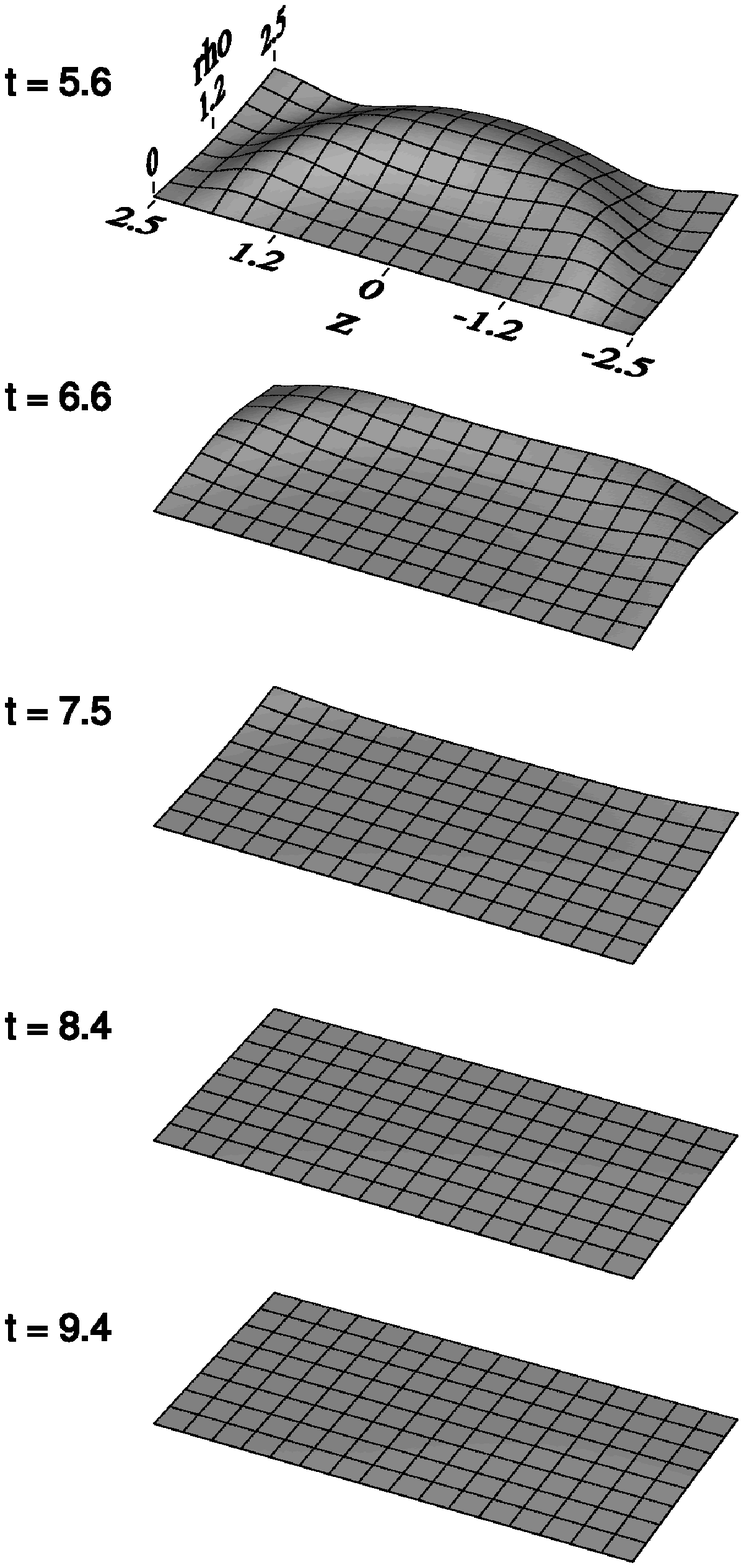}}
\caption{The metric variable $\bs$ from two sample Brill wave evolutions, 
         demonstrating the effectiveness of dissipation. Both simulations
         were run with identical parameters, except the 
         simulation displayed
         on the left was run without dissipation,
         while that shown on the right 
         was run with a dissipation parameter $\epsilon=0.5$ 
         (see Appendix \ref{app:solver}).
         The simulation without dissipation crashed at around $t=10$,
         while the run with dissipation was stopped after several light
         crossing times, without showing any signs of instability.
         (Note: the computational domain used for these evolutions was 
          $\rho=[0,10]$, and $z=[-10,10]$; in the plots above we only show
          a subset of this domain. Also, the grid-lines drawn are for visual
          aid only; the size of the mesh visible in each frame is
          $64\times 128$.)
         }
\label{fig:diss}
\end{figure}

We add Kreiss-Oliger dissipation~\cite{ko} to the evolution of equations
of $\Phi,\Pi,\bs$ and $\bo$ (in addition to $\psi$ during
partially constrained evolution), as described in Appendix \ref{app:solver}.
To demonstrate that this is {\em essential} for the stability
of our numerical scheme, we compare in Fig. \ref{fig:diss} 
the evolution of $\bs$ from simulations without and with dissipation,
but otherwise identical.

The elliptic equations (\ref{eq:cons_alpha}-\ref{eq:cons_betaz}) 
are solved using Brandt's FAS multigrid (MG)
algorithm \cite{brandt,brandt2}, described in some detail in Appendix 
\ref{app:solver}. There are no explicit time derivatives of functions
in these equations, and we discretize them at a single time level $n$ ({\it i.e.} we 
do not apply the Crank-Nicholson averaging scheme for the elliptics).
We use either a fully constrained evolution, solving for
$\alpha,\beta^\rho,\beta^z$ and $\psi$ using the constraint
equations and slicing condition, or a partially constrained 
evolution 
where instead of using the Hamiltonian constraint to update $\psi$, 
we use the evolution equation ($\ref{eq:evol_psi}$).
Partially constrained evolution has proven to be useful due to the 
occasional failure of the MG solver in the strong-field regime 
({\em i.e.} close to black hole formation).  Use of the evolution 
equation for $\psi$ (rather than the Hamiltonian constraint) circumvents 
this problem in many instances; however, in certain Brill-wave dominated 
spacetimes, free evolution of $\psi$ is not sufficient to ensure 
convergence of the MG process.
We are currently working to make the MG solver more robust in 
these situations.

The code is written in a combination of RNPL (Rapid Numerical 
Prototyping Language \cite{RNPL}) and Fortran 77. The hyperbolic
equations are implemented in RNPL, which employs a point-wise
Newton-Gauss-Seidel iterative relaxation scheme to solve these
equations, while the MG solver is implemented in Fortran (see
Appendix \ref{app:solver} for more MG details). A pseudo-code
description of the time-stepping algorithm used is as follows:

\begin{center}
\begin{verbatim}
a) As an initial guess to the solution at time t+dt, copy variables 
   from t to t+dt

b) repeat until (residual norm < tolerance):
      1: perform 1 Newton-Gauss-Seidel relaxation sweep of the 
         evolution equations, solving for the unknowns at time t+dt
      2: perform 1 MG vcycle on the set of elliptic equations at 
         time t+dt
   end repeat
\end{verbatim}
\end{center}
For the residual norm used to terminate the iteration we use the 
the infinity norm of the residuals of all updated unknowns.

\section{Tests}
\label{sec:test}
In this section, we describe some of the tests we have performed
to check that we are solving the correct set of equations.
The first test consists of checking the equations against those 
derived with a computer algebra
system~\cite{maple}. By inputting the metric and coordinate conditions,
the  computer derived equations can then be subtracted from our
equations and simplified. By finding that the differences simplify to 
0, we can conclude that two sets of equations agree.

For diagnostic purposes and as tests of the equations and of their 
discretization, 
we compute several quantities during the numerical evolution.   
The first is the ADM mass \cite{MTW}
\begin{equation}\label{MADM}
M_{ADM}=\frac{1}{16\pi} \lim_{r \rightarrow \infty}
\int \left(H^a_{\ b;a}-H^a_{\ a;b}\right) N^b d A,
\end{equation}
where the integral is evaluated on a flat 3-space,
{\it i.e.} with metric $ds^2=d\rho^2+dz^2+\rho^2 d\phi^2$.  The spatial
3-metric $H_{ab}$ is that from our curved space solution, but has its indices
raised and lowered with the flat metric. Integrating
around the boundaries of our numerical grid, the normal vectors $\vec{N}$
are $\pm \partial/\partial z$ and $\partial/\partial \rho$.
After some algebra, the ADM mass becomes
\beq
M_{\rm ADM} 
  & = & \;\frac{1}{2} \int_{z_{\rm max}}
          \rho \psi^4 \left[ - \frac{\psi_{,z}}{\psi}
                             - e^{2\rho\bs}
                                \Bigl(   \frac{\psi_{,z}}{\psi}
                                       + \frac{1}{2} \bigl(\rho\bs\bigr)_{,z}
                                \Bigr)
                     \right] d\rho \cr
  &   &-\frac{1}{2} \int_{z_{\rm min}}
          \rho \psi^4 \left[ - \frac{\psi_{,z}}{\psi}
                             - e^{2\sigma}
                                \Bigl(   \frac{\psi_{,z}}{\psi}
                                       + \frac{1}{2} \bigl(\rho\bs\bigr)_{,z}
                                \Bigr)
                     \right] d\rho \cr 
  &  & +\frac{1}{2} \int_{\rho_{\rm max}}  
          \rho \psi^4 \left[ - \frac{\psi_{,\rho}}{\psi}
                             - e^{2\sigma}
                                \Bigl(
                                         \frac{\psi_{,\rho}}{\psi}
                                       + \frac{1}{2} \bigl(\rho\bs\bigr)_{,\rho}
                                       + \frac{1}{4\rho}
                                \Bigr)
                             + \frac{1}{4\rho} 
                     \right] dz.
\eeq

A second set of quantities we calculate are
the $\ell_2$-norms of the residuals of
the evolution equations for the extrinsic curvature~(\ref{eq:k_AB_dot}), 
which we denote $E(\dot{K}_{AB})$. Because we do not directly evolve individual 
components of the extrinsic curvature, these residuals will not be zero; 
however, they {\em should} converge to zero in the limits as the 
discretization scale $h\rightarrow 0$,
and the outer boundary positions $\rho_{\rm max},z_{\rm max},-z_{\rm min} \rightarrow
\infty$.  Note that we include these last conditions because it is only in the 
limit $r\rightarrow \infty$ that our
outer boundary conditions are fully consistent with asymptotic 
flatness.

The convergence properties of our code are measured by computing the
convergence factor, $Q_u$, associated with a given variable, $u$,
obtained on grids with resolution $h$, $2h$ and $4h$ via
\be
Q_u = \frac{    || u_{4h} - u_{2h} ||_2 }
           {    || u_{2h} - u_{h}  ||_2 }.
\ee
In particular, for the case of $O(h^2)$ (second order)
convergence, we expect $Q_u \to 4$ as
$h\to 0$.

\begin{figure}
\epsfxsize=16cm
\centerline{\epsffile{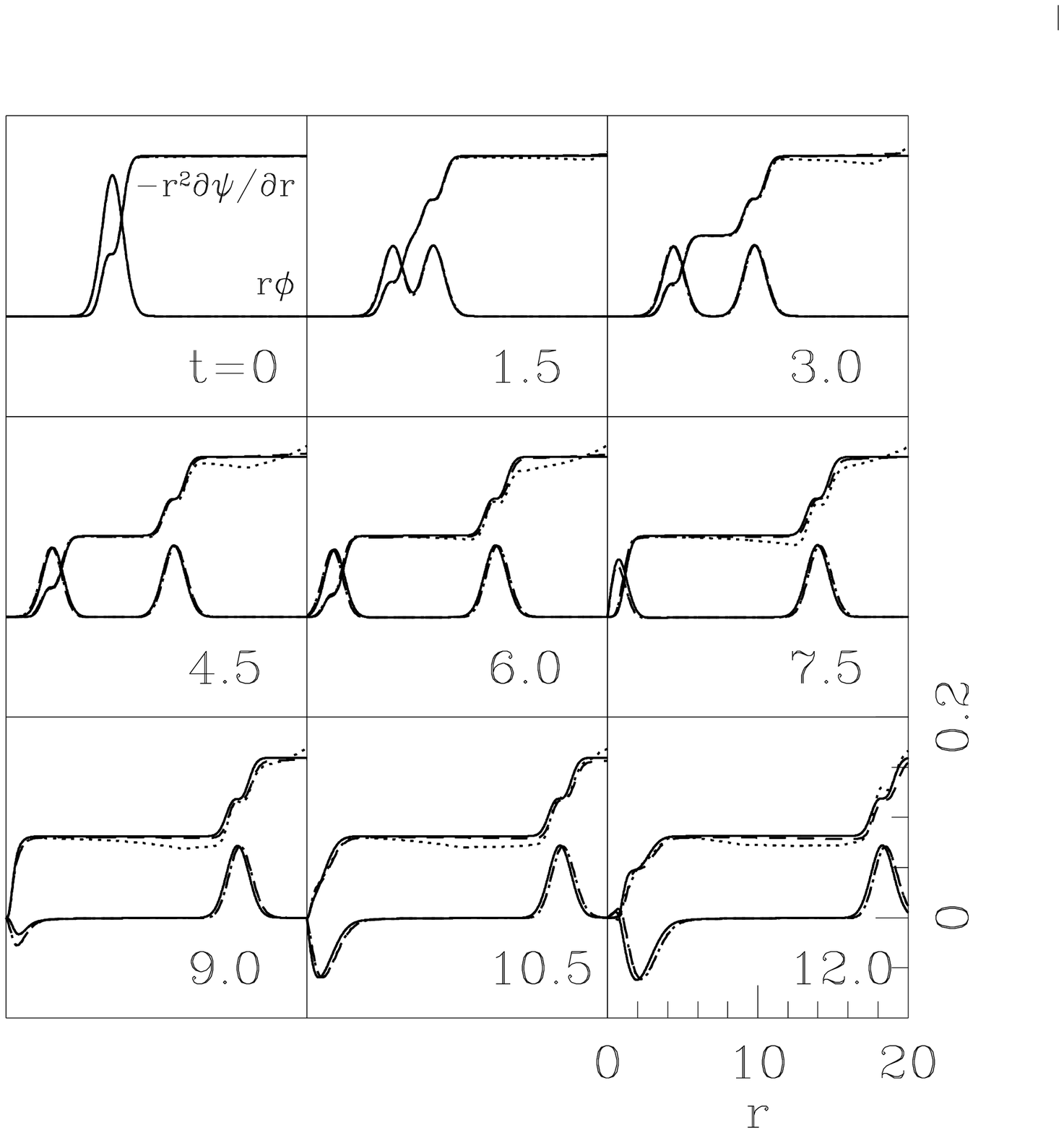}}
\caption{Tests of spherically symmetric  scalar collapse. The sequence of
         frames are from the evolution of an initial pulse in $\Phi$ of 
         the form~(\ref{G_X}) with
         $A_\Phi = 0.02$,
         $\epsilon_\Phi = 1$,
         $R_\Phi = 7.0$,
         $\Delta_\Phi = 1.0$,
         $(\rho_\Phi, z_\Phi)=(0,0)$.
         Shown are the functions $r \Phi$ and $-r^2 \partial \psi/\partial r$.
         The output of the explicitly spherically symmetric code
         (with $2^{10}$ radial grid points)
         is shown with solid lines, while the output of the axisymmetric code
         ($N_\rho = N_z/2 = 2^8$) is shown with dashed 
         (a $(\rho =0, z>0)$ slice) and dotted lines (a $z=0$ slice).}
\label{fig:spherical}
\end{figure}

The first set of tests we present here are comparisons of $\Phi$ and $\psi$ from the evolution of 
spherically symmetric initial data to 
the corresponding functions computed by a 1D spherically symmetric
code, the details of which are presented
in Appendix~\ref{app:spherical}. In general, the results from the 
two codes are in good
agreement. A sample comparison is illustrated in Fig.~\ref{fig:spherical} which
shows the scalar field obtained with the 1D code as well as
two radial slices of the corresponding solution calculated using the 
axisymmetric code. Note, however, that we do not expect {\em exact} agreement 
in the
limit $h\rightarrow 0$ for a fixed outer boundary location, as the 
``rectangular''
boundaries of the axisymmetric code are, in general, incompatible with precise 
spherical symmetry.

\begin{figure}
\epsfxsize=8.5cm
\centerline{\epsffile{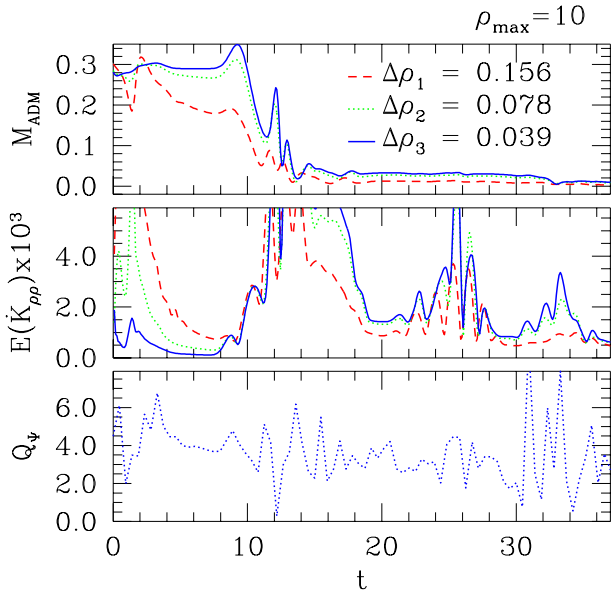}}
\caption{Tests of Brill collapse
         using a initial pulse profile for $\bs(0,\rho,z)$ of the form 
         (\ref{G_X}) with
         $A_{\bs} = -3.0$, $R_{\bs} = 0$, $\Delta_{\bs} = 1$,
         $\epsilon_{\bs} = 1$, and $(\rho_{\bs},z_{\bs})=(0,0)$. 
         The evolution shown here corresponds to 
         four crossing times and $\rho_{\rm max}
         = z_{\rm max} = 10$. 
         The top frame shows the calculated ADM mass $M_{\rm ADM}$.
         As the resolution increases, so does the level of
         mass conservation at early times, before energy
         has reached the outer boundary. The middle frame shows
         $E(\dot{K}_{\rho\rho})$,
         the $\ell_2$-norm of the residual of the $\rho\rho$ component of
         the evolution equation for the extrinsic curvature~(\ref{eq:k_AB_dot}).
         At early times, before energy reaches the outer boundary,
         the residual decreases as the resolution
         increases. The bottom frame shows the convergence factor
         computed for the field $\psi$.
         }
\label{fig:brill_10}
\end{figure}

\begin{figure}
\epsfxsize=8.5cm
\centerline{\epsffile{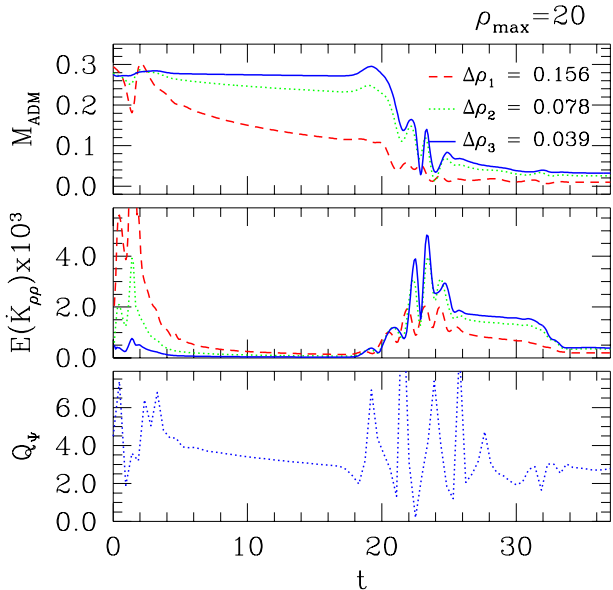}}
\caption{Tests of Brill collapse with the same initial data and grid
         resolutions as those in Fig.~\ref{fig:brill_10} above, but with
         $\rho_{\rm max}= z_{\rm max} = 20$, and for only
         two crossing times. Notice the improvement
         in the behavior of the residual and mass aspect when
         energy reaches the outer boundary, as compared to the
         $\rho_{\rm max}= z_{\rm max} = 10$ case above.
         }
\label{fig:brill_20}
\end{figure}

\begin{figure}
\epsfxsize=8.5cm
\centerline{\epsffile{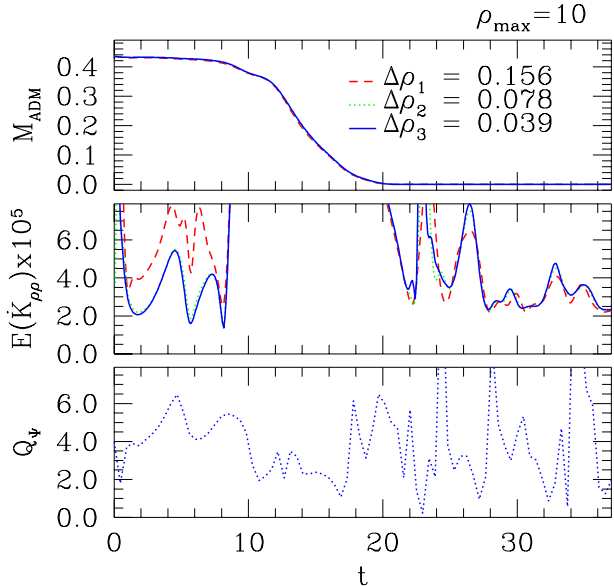}}
\caption{Tests of an oblate, scalar pulse evolution
         with $A_\Phi = 0.15$, $R_\Phi = 0$, $\Delta_\Phi = 3$,
         $\epsilon_\Phi = 3$, and $(\rho_\Phi,z_\Phi)=(0,0)$. The
         tests are shown for four crossing times with $\rho_{\rm max}
         = z_{\rm max} = 10$. The results are similar to the tests
         for the Brill wave shown in Fig.~\ref{fig:brill_10} above,
         though notice that the scale of the residual $E(\dot{K}_{\rho\rho})$ 
         is about 2 orders of magnitude smaller than that of the Brill wave case. 
         }
\label{fig:oblate_10}
\end{figure}

\begin{figure}
\epsfxsize=8.5cm
\centerline{\epsffile{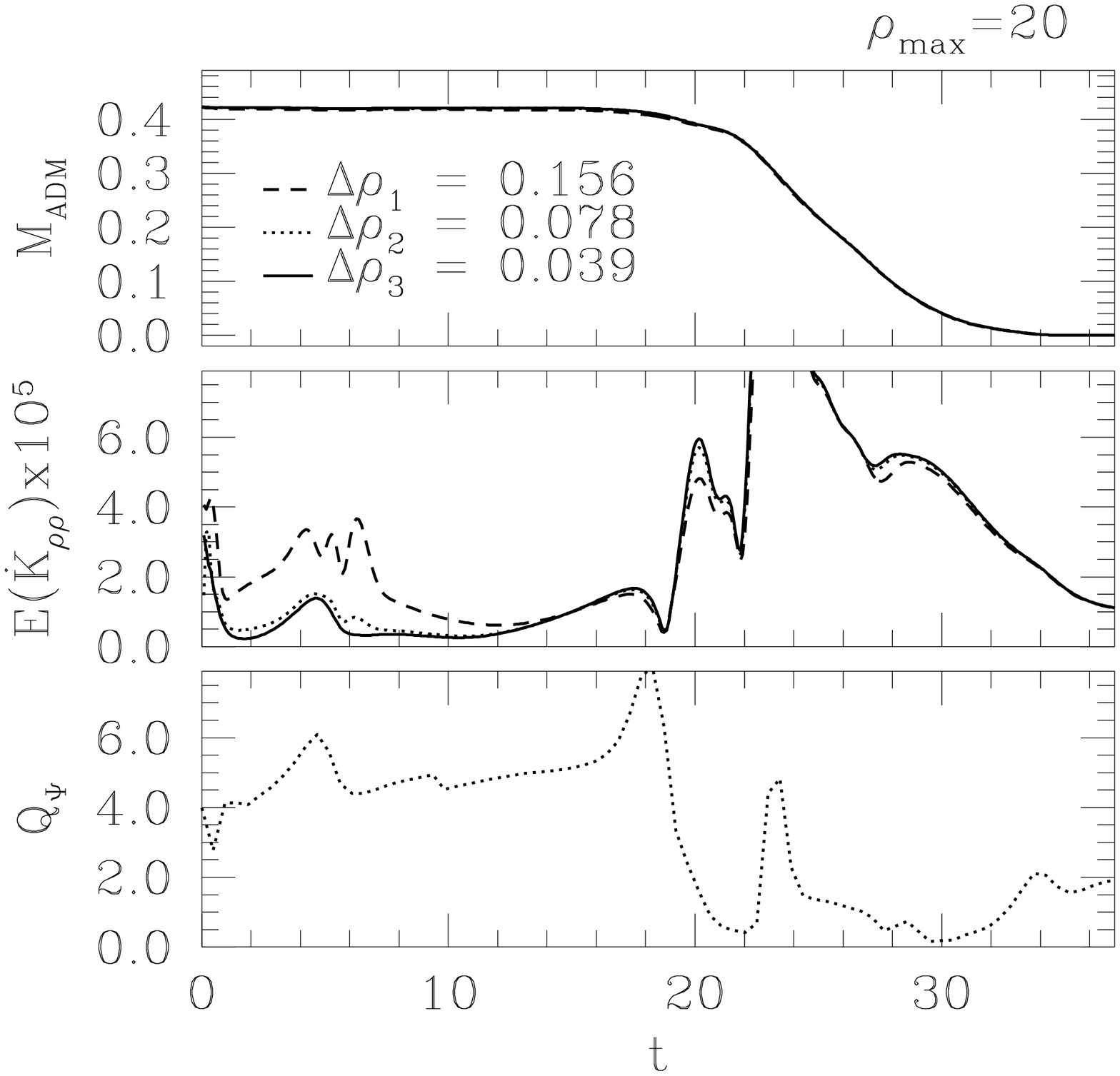}}
\caption{Tests of an oblate, scalar pulse evolution with
         the same initial data and grid
         resolutions as those in Fig.~\ref{fig:oblate_10} above, but with
         $\rho_{\rm max}= z_{\rm max} = 20$, and for 
         two crossing times. Again, as with the Brill wave example, it is
         evident that there are two factors that contribute to a non-zero
         residual $E(\dot{K}_{\rho\rho})$---the closeness of the outer
         boundary, and the discretization scale $h$.
         }
\label{fig:oblate_20}
\end{figure}

In the second series of tests,
we examine evolutions of Brill waves and
non-spherical scalar pulses. 
Figs.~\ref{fig:brill_10}--\ref{fig:oblate_20} show results 
from two typical initial data sets, each computed using two 
distinct outer boundary positions.
Each figure plots (a) the ADM mass $M_{\rm ADM}$,
(b) the $\ell_2$-norm of the residual of the $\rho\rho$ component of 
the evolution equation for the extrinsic curvature $E(\dot{K}_{\rho\rho})$,
and, (c) the convergence factor $Q_\psi$ of $\psi$, as functions of time
(the convergence factor $Q$ for other functions exhibit similar behavior
as $Q_\psi$, and so for brevity we do not show them). 
Here, one expects to see an improvement of the results---namely
trends toward mass conservation early on, a zero residual, and a 
convergence factor of 
4---in the limits $h \to 0$ and $(\rho_{\rm max}, z_{\rm max}) \to \infty$.
After energy has reached the outer boundary,
and to a lesser extent before (as is evident in the scalar field
example in Figs.~\ref{fig:oblate_10} and ~\ref{fig:oblate_20}),
we fail to get consistency with the evolution equation~(\ref{eq:k_AB_dot})
as $h \to 0$, for {\em fixed} $(\rho_{\rm max}, z_{\rm max})$.
This is a measure of the inaccuracy of our outer boundary 
conditions~(\ref{eq:obc_alpha}-\ref{eq:obc_betarho}); though the trends suggest
that we {\em do} achieve consistency in the limit 
$(\rho_{\rm max}, z_{\rm max}) \to \infty$. 

\begin{figure}
\centerline{\epsfxsize=7cm \epsffile{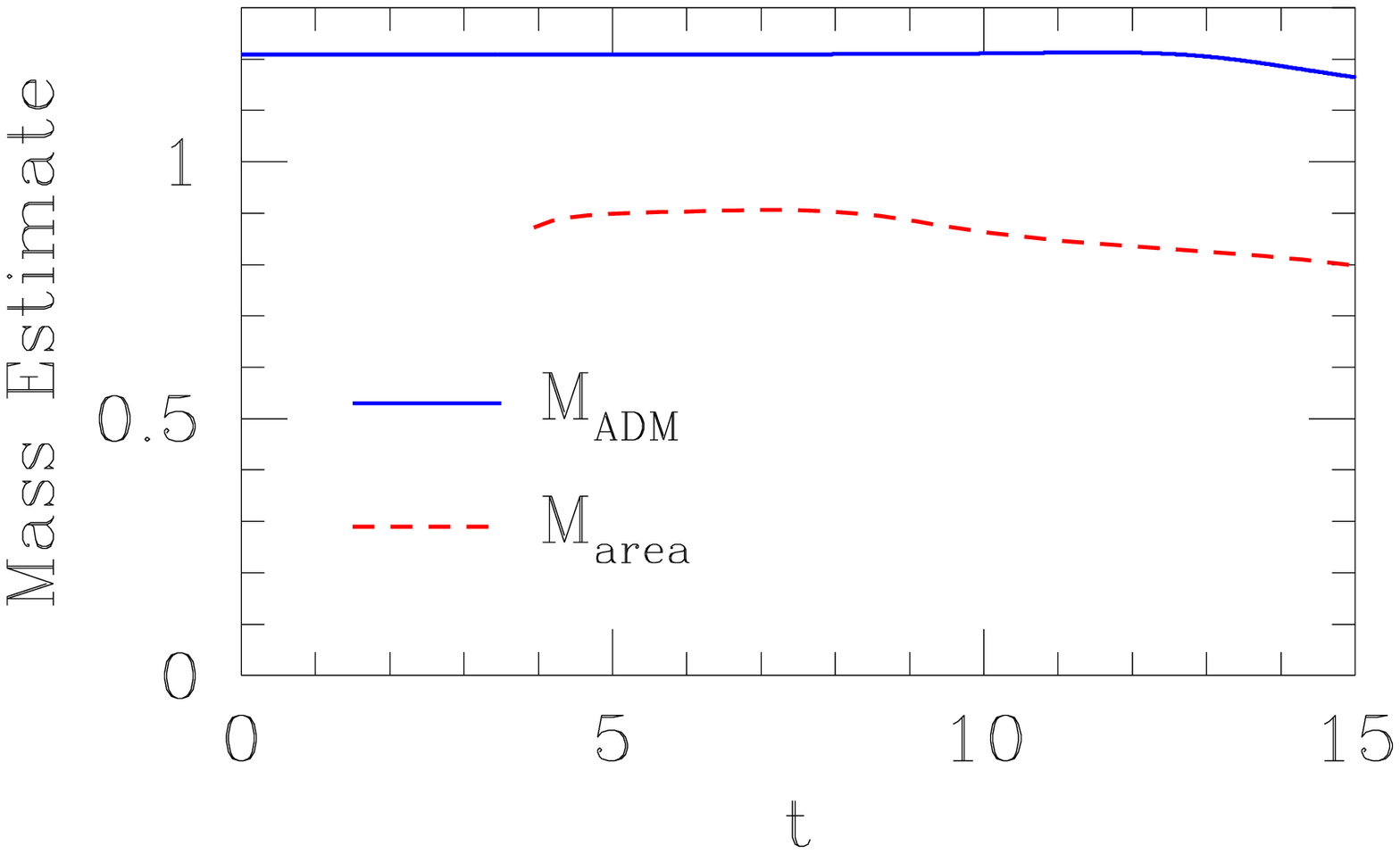}
             \epsfxsize=7cm \epsffile{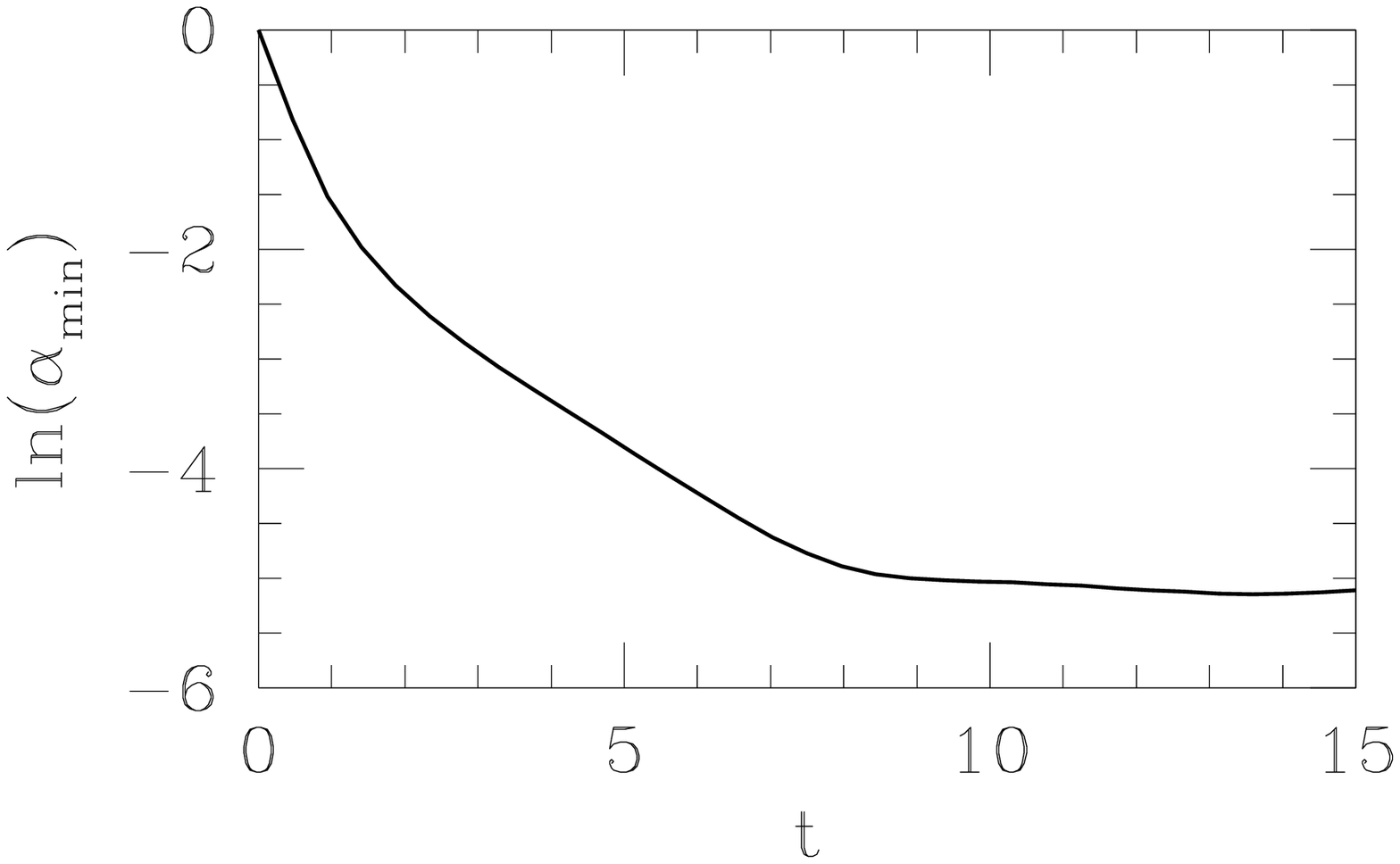}}
\caption{The ADM mass estimate $M_{\rm ADM}$ of the spacetime and 
area-mass estimate $M_{\rm area}=\sqrt{A/16}$ of the black hole (left), 
where $A$ is the area of the apparent horizon, and
the natural logarithm of the minimum value of lapse (right)
during a scalar field collapse simulation. The initial 
scalar field profile, $\Phi(0,\rho,z)$ is of the form
(\ref{G_X}) with
$A_\Phi=0.35, R_\Phi=0, \Delta_\Phi=1, \epsilon_\Phi=1$ and
$(\rho_\Phi,z_\Phi)=(0,0)$. The outer boundary is at
$\rho_{\rm max}=z_{\rm max}=-z_{\rm min}=10$, and the size of the numerical
grid is $256\times 512$. An apparent horizon was first detected
at $t \approx 4$, hence the $M_{\rm area}$ curve only starts then.
At intermediate times (between roughly $t=2$ and $t=7$)
we see an exponential ``collapse'' of the lapse 
(the minimum of which is at $(\rho,z)=(0,0)$);
at later times this behavior ceases in the simulation, as 
a consequence of increasingly poor resolution in the vicinity of the
black hole caused by ``grid-stretching''---see Fig.
\ref{fig:bh_psi} below. This also adversely affects the accuracy of the area-mass
estimate (it actually begins to {\em decrease} at late times)
as the coordinate region occupied by the AH shrinks.
}
\label{fig:bh_plots}
\end{figure}

\begin{figure}
\centerline{\epsfxsize=15cm \epsffile{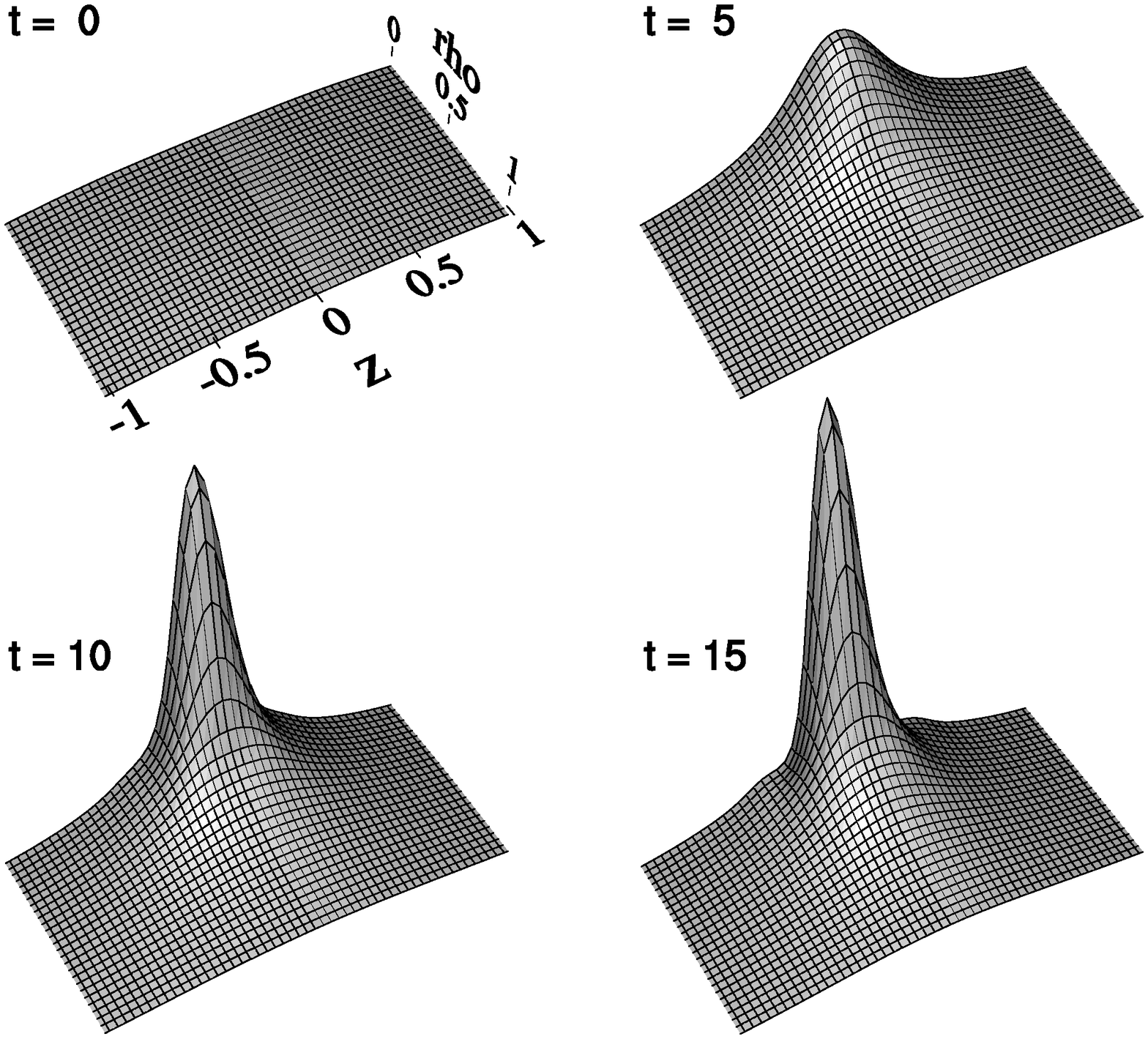} }
\caption{The conformal factor $\psi$ near the origin, 
at four different times from the simulation described in 
Fig. \ref{fig:bh_plots}. The height of the surfaces represent
the magnitude of $\psi$, and the scale of each image is the
same. The smallest value of $\psi$ shown in each frame is
$\approx 1.5$, while at $t=15$, $\psi$ reaches a maximum of 
$\approx 5.8$ at the origin. The $2$--metric has the
form $\psi^4(d\rho^2+dz^2)$, hence the larger $\psi$ is,
the larger the physical area represented by a given
coordinate cell (the lines drawn here {\em are} coincident with
the actual grid-lines of the simulation). This implies that the effective
resolution of a given coordinate patch {\em decreases} as $\psi$ increases.
The net result is that as gravitational collapse proceeds,
central features of the solution become very poorly resolved 
within the grid, adversely affecting the accuracy of the solution.
This is quite evident at $t=15$, where 
noticeable asymmetries have developed in $\psi$ (recall
that this is collapse from spherically symmetric initial data).
}
\label{fig:bh_psi}
\end{figure}

Finally, we show some results of a simulation of black hole formation
from the collapse of a spherically symmetric distribution of scalar field
energy. Again, by looking at spherically symmetric collapse we can
compare with the 1D code (obtaining the same level of agreement
as seen in the example in Fig. \ref{fig:spherical}). 
However, here we want to show the behavior
of our coordinate system (in particular the maximal slicing) in the 
strong-field regime, which demonstrates
the need to incorporate black hole excision techniques and/or adaptive
mesh refinement (AMR) before attempting any serious investigation 
of physics with this code. Fig. \ref{fig:bh_plots} shows plots of 
the ADM mass estimate (\ref{MADM}), an estimate of the black hole
mass $M_{\rm area} \equiv \sqrt{A/16\pi}$, where $A$ is the area of
the apparent horizon, and the minimum value of the lapse as a 
function of time from the simulation.
Fig. \ref{fig:bh_psi} shows the conformal factor
$\psi$ at several times, in the central region of the grid.

Maximal slicing is considered {\em singularity avoiding}, because
as the singularity is approached in a collapse scenario, the 
lapse $\alpha$ tends to 0, as demonstrated in Fig. 
\ref{fig:bh_plots}. This effectively freezes the evolution inside
the black hole, though it causes a severe distortion in the 
$t={\rm const.}$ slices as one moves away from the black hole. This
particular coordinate pathology is evident in Fig. \ref{fig:bh_psi}. 
Recall from the $2$--metric (\ref{HAB_def}) that $\psi^2$ determines
proper length scales in the $\rho$ and $z$ directions; thus
the rapid growth with time of $\psi$ shown in Fig. \ref{fig:bh_psi} means
that a given coordinate area represents increasing proper area. 
Furthermore, the increase in magnitude of $\psi$ in the
strong-field regime (which happens even when black holes do not form,
and in non-spherical scalar field and Brill wave evolution,
though not to the same extent as shown in Fig. \ref{fig:bh_psi})
implies that our effective numerical resolution decreases in those
regions, as some feature of the solution with a given characteristic
size will span less of the coordinate grid. 
Thus, in the end, even though maximal slicing may prevent us from
reaching a physical singularity, the ``grid-stretching'' effect 
is just as disastrous
for the numerical code, preventing any long-term simulation of 
black hole spacetimes. For these reasons
we will add black hole excision techniques and AMR before exploring
physics with this code; our efforts in this regard are
well underway, and will be described elsewhere. 

\section{Conclusion}
\label{sec:conclusion}

We have described a $(2+1)+1$ gravitational evolution model
which evolves axisymmetric configurations of gravitational radiation
and/or a scalar field. A thorough battery of tests confirms that the
correct equations are being solved. In particular, we have provided 
evidence
that the code is second-order convergent, consistent and conserves mass 
in the limit where the outer boundary position goes to infinity.

The unigrid code described here is the first step towards our
long-term goal of studying a range of interesting theoretical and
astrophysical phenomena in axisymmetry. These include gravitational
collapse of various matter sources and gravitational waves, the
corresponding critical phenomena at the threshold of black hole formation, 
head-on black hole collisions and accretion disks. To this end, we need to 
include support for 
angular momentum and additional matter fields in the code, as well
as to add additional computational and mathematical infrastructure---adaptive
mesh refinement, black hole excision and the capability of running
in parallel on a network of machines. All of these projects are
under development, and results will be published as they become
available.

Another goal of this project is to provide a platform from which to develop 
computational technology for 3D work. In particular, we see development
of AMR in axisymmetry as a precursor to its incorporation in 3D
calculations.
Likewise, accurate and stable treatment of boundary conditions presents a
continual challenge in numerical relativity,
and it is possible that we can develop an effective treatment of boundaries
in axisymmetry that will generalize
to the 3D case.

\section*{Acknowledgments}
\label{sec:ack}
The authors would like to thank David Garfinkle for helpful discussions.
MWC would like to acknowledge financial support from 
NSERC and NSF PHY9722068. FP would like to acknowledge
NSERC, The Izaak Walton Killam Fund
and Caltech's Richard Chase Tolman Fund for their financial support. 
EWH and SLL would like to acknowledge
the support of NSF Grant PHY-9900644. SLL acknowledges support
from grant NSF PHY-0139980 as well as
the financial support of Southampton College.
EWH also acknowledges the support of NSF grant PHY-0139782.
The majority of the simulations described here were performed on 
UBC's {\bf vn}
cluster (supported by CFI and BCKDF), and the {\bf MACI} cluster 
at the University of Calgary (supported by CFI and ASRA).
\appendix

\section{The Elliptic Solver and Finite Difference Approximations}
\label{app:solver}

In this appendix we briefly mention some aspects of our multigrid (MG)
routine, and list the set of finite difference approximations that we use.

The constraint equations~(\ref{eq:cons_alpha}-\ref{eq:cons_betaz})
are four elliptic equations which, for a fully constrained system,
must be solved on every time slice ({\it i.e.} spatial hypersurface). As such,
it can be expected that the time taken by a given evolution will
be dominated by the elliptic solver and hence we look for the fastest
possible solver.

Currently, multigrid methods are among the most efficient elliptic
solvers available, and here we have implemented a standard Full 
Approximation Storage
(FAS) multigrid method with V--cycling (see \cite{brandt,brandt2}) to 
solve the four nonlinear equations simultaneously.
(When using the evolution equation for $\psi$ in a partially constrained
evolution, we use the same multigrid routine described here,
except we only solve for the three quantities $\alpha,\beta^\rho$
and $\beta^z$ during the V-cycle; $\psi$ is then simply considered
another ``source function''.)

\begin{figure}
\epsfxsize=5cm
\centerline{\epsffile{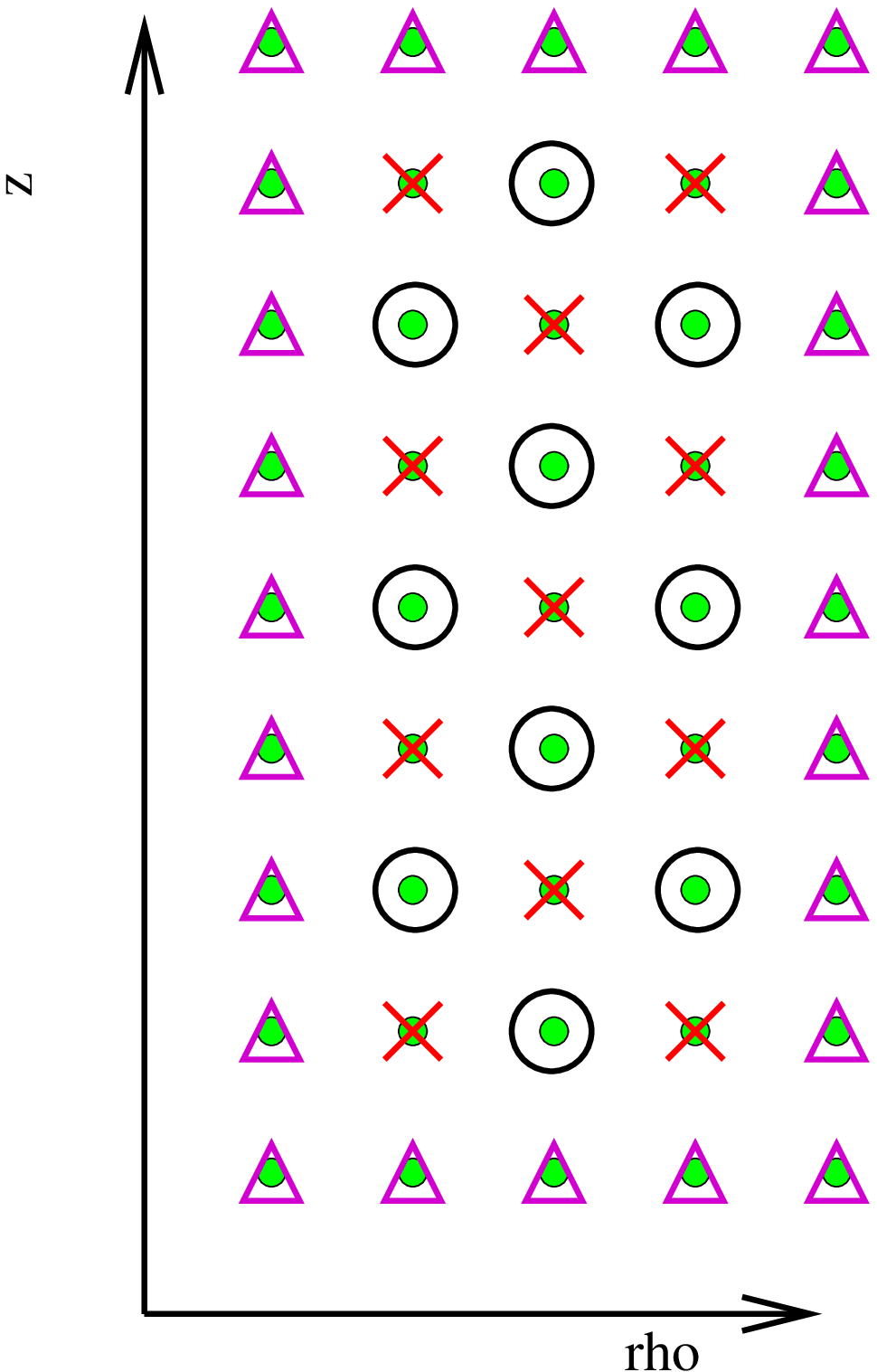}}
\caption{Diagram illustrating the order in which relaxation occurs.
         Shown here is a $5\times 9$ grid spanning $\rho \times z$
         with the grid points represented by filled circles.
         The Xs denote ``red'' (interior) points which are visited
         first. The Os denote  ``black'' (interior) points which
         are visited next. Finally, the triangles denote boundary
         points which are visited last.
         }
\label{fig:redblack}
\end{figure}

A key component of the MG solver is the relaxation routine that is 
designed to {\em smooth} the residuals associated with the 
discretized elliptic equations.
We use point-wise Newton-Gauss-Seidel relaxation with red-black ordering
(see Fig.~\ref{fig:redblack}),
simultaneously updating
all four quantities $(\alpha,\beta^\rho,\beta^z,\psi)$ at
each grid-point during a relaxation sweep.
In addition to its use for the standard pre-coarse-grid-correction (pre-CGC)
and post-CGC smoothing sweeps, the relaxation routine is also used to
{\em solve} problems on the coarsest grid.
We use half-weighted restriction to transfer fields from fine to coarse
grids and (bi)linear interpolation for coarse to fine transfers.
We generally use 3 pre-CGC and 3 post-CGC sweeps per $V$-cycle, and likewise
normally use a single $V$-cycle per Crank-Nicholson iteration.

One complicating factor here is the treatment of the boundary
conditions Eqs.~(\ref{eq:bc_alpha}-\ref{eq:bc_betarho}) and
Eqs.~(\ref{eq:obc_alpha}-\ref{eq:obc_betarho}).
In accordance with general multi-grid practice, we view the boundary 
conditions as logically and operationally distinct from the interior 
equation equation.
The outer boundary conditions can generally be expressed as
\be
r\,X \approx {\rm constant} \label{ap:bc}
\ee
where $X \in \left\{ 1-\alpha, 1-\psi, \beta^z, \beta^\rho \right\}$. Taking
the derivative of (\ref{ap:bc}) with respect to $r$, 
we arrive at the differential form that is applied on the outer 
boundaries of the computational domain ($\rho=\rho_{\rm max}, z=z_{\rm max},
z=-z_{\rm max}$):
\be
X - \rho X_{,\,\rho} - z X_{,z} = 0.
\ee
On the $z$-axis ($\rho=0$), the conditions
Eqs.~(\ref{eq:bc_alpha}-\ref{eq:bc_betarho}) take one of the following two 
forms:
\be
A_{i,\rho} = 0
\label{eq:qfit}
\ee
or
\be
A_i = 0
\label{eq:zfit}
\ee
Equations of the former form are discretized using an $O(h^2)$ backwards 
difference approximation to the $\rho$ derivative.

The interior and boundary differential equations 
are solved in tandem via the multigrid approach:
\begin{enumerate}
\item The residual is smoothed using some number of relaxation sweeps.
      For the interior, Eqs.~(\ref{eq:cons_alpha}-\ref{eq:cons_betarho}) 
      are relaxed using red-black ordering
      as discussed above. After each call of this relaxation routine, a second
      routine that ``relaxes'' the boundary points is called~\cite{boundary}.
\item For quantities restricted from a fine to a coarse grid, discrete forms
      of (\ref{eq:qfit}) and
      (\ref{eq:zfit}) are applied during the
      $V$-cycle. At the other boundaries, straightforward injection is used.
\end{enumerate}
The key idea here is to ensure 
that the boundary relaxation process does not substantially impact
the smoothness of the interior residuals, because it is only for smooth 
residuals that a coarsened version of a fine-grid problem can 
sensibly be posed.

Finally, in Table \ref{table:diff}, we show all of the difference
operators we use to convert the differential 
equations listed in \ref{sec:coords_and_vars} to finite difference form,
using the Crank-Nicholson scheme described in Section \ref{sec:implementation}.
In addition, as discussed in Section \ref{sec:implementation}, 
we use Kreiss-Oliger dissipation~\cite{ko} to maintain smoothness in the evolved fields. 
Specifically, we add the Kreiss-Oliger filter to discretized 
evolution equations
\be
\Delta_t A^n = \mu_t f^n\left(\dots\right)
\ee
by replacing the Crank-Nicholson time difference operator
$\Delta_t$ with $\Delta^\epsilon_t$:
\be
\Delta^\epsilon_t A^n = \mu_t f^n\left(\dots\right)
\ee
Empirically, we find that a value of $\epsilon = 0.5$
generally keeps our fields acceptably smooth.

\begin{table}
  \begin{center}
     \begin{tabular}{l l l}
     \hline
     Operator & Definition & Expansion\\
     \hline
$\Delta_x A_i$         
   &
   $\left( A_{i+1} - A_{i-1} \right) / \left( 2 h \right)$
   &
   $\left. A_{,x}\right|_i + O\left(h^2\right)$
\\
$\Delta_x^f A_i$
   &
   $\left( -3A_{i} + 4A_{i+1}  -  A_{i+2}\right) / \left( 2 h \right)$
   &
   $\left. A_{,x}\right|_i + O\left(h^2\right)$
\\
$\Delta_x^b A_i$
   &
   $\left(  3A_{i} - 4A_{i-1}  +  A_{i-2}\right) / \left( 2 h \right)$
   &
   $\left. A_{,x}\right|_i + O\left(h^2\right)$
\\
$\Delta_{xx} A_i$
   &
   $\left( A_{i+1} -2 A_i + A_{i-1} \right) / \left( h^2 \right)$
   &
   $\left. A_{,xx} \right|_i + O\left(h^2\right)$
\\
$\Delta_{xxxx} A_i$
   &
   $\Delta_{xx} \left( \Delta_{xx} A_i \right)=\left(A_{i+2} -4 A_{i+1} + 6 A_i
   -4 A_{i-1} + A_{i-2} \right)/h^4$
   &
   $\left. A_{,xxxx}\right|_i + O\left(h^2\right)$
\\
$\Delta_{x^2} A_i$
   &
   $\left( A_{i+1} - A_{i-1} \right) / \left( x_{i+1}^2 -x_{i-1}^2 \right)$
   &
   $\left. A_{,xx}\right|_i + O\left(h^2\right)$
\\
$\Delta_{x} (A_i/x)$
   &
   $2 \left( x_{i-1} A_{i+1} - x_{i+1} A_{i-1} \right) / (x_{i+1}^2 - x_{i-1}^2 ) / x_i$
   &
   $\left. (A/x)_{,x} \right|_i + O\left(h^2\right)$
\\
$\Delta_x[(\Delta_{x} A_i)/x]$
   &
   $16 \left[x_{i-1/2} A_{i+1} - 2 x_i A_i + x_{i+1/2} A_{i-1} \right] / ( x_{i+1}^2 - x_{i-1}^2 )^2$
   &
   $\left. (A_{,x}/x)_{,x} \right|_i + O\left(h^2\right)$
\\
$\mu_t A^n$
   &
   $\left( A^{n+1} + A^{n} \right) / 2        $                            
   &
   $\left. A\right|^{n+1/2} + O\left(\lambda^2h^2\right)$
\\
$\Delta_t A^n$
   &
   $\left( A^{n+1} - A^{n} \right) / \left(\lambda h\right)$               
   &
   $\left. A_{,t}\right|^{n+1/2} + O\left(\lambda^2h^2\right)$
\\
$\Delta_t^\epsilon A^n$
   &
   $\left[ \Delta_t + \epsilon h^3 / \left(16\lambda\right) \Delta_{xxxx} \right] A^n_i$
   &
   $\left. A_{,t} \right|^{n+1/2} + O\left(\lambda^2h^2\right)$
\\
     \hline
     \hline
     \end{tabular}
  \end{center}
  \caption{Finite difference operators and their correspondence to 
           differential operators. Here,
           $A^n_i$ is an arbitrary grid function defined via
           $A^n_i \equiv A(x_{\rm min} + (i-1)h,t_{\rm min} + 
                 (n-1)(\lambda h))$,
           where $h$ and $\lambda h$ are the 
           spatial and temporal grid spacings, respectively. 
           $x$ denotes either of
           the two spatial coordinates $\rho$ or $z$, with the dependence
           of $A^n_i$ on the other suppressed.
           The parameter $\epsilon$ represents a user-specifiable ``amount'' of
           Kreiss-Oliger dissipation.
           }
  \label{table:diff}
\end{table}

\section{The Spherically Symmetric Model}
\label{app:spherical}
One simple test of the code compares the results for spherically
symmetric initial data with the output of a code which explicitly
assumes spherical symmetry. Here we present the equations
for this 1D code.  The spacetime metric is:
\be
ds^2 = -(\alpha^2 + \psi^4\beta^2)dt^2 + 2\psi^4\beta dt dr + \psi^4 \left(   dr^2 + r^2 d\Omega^2 \right),
\ee
where $\alpha$, $\beta$ and $\psi$, are functions of $r$ and $t$, 
$d\Omega^2$ is the line element on the unit 2-sphere, and $\beta$ is 
the radial component of the shift vector (i.e. $\beta^i = (\beta,0,0)$).
Adopting maximal slicing to facilitate direct comparison to the 
axisymmetric code, we have 
\be
	K^i{}_j = {\rm diag}(K^r{}_r(r,t),0,0).
\ee
Then a sufficient set of equations for the coupled Einstein-massless-scalar 
system is~\cite{mattsthesis}
\beq
\psi'' + \frac{2\psi'}{r} + 2 \pi \left[ \Phi^2 + \Pi^2 \right] \psi + \frac{3}{16} \left( K^r_r\right)^2 \psi^5
& = &  0
\label{eq:ss_psi}
\\
\left( K^r{}_r \right)' + 3 \frac{\left(r\psi^2\right)'}{r\psi^2}  K^r{}_r + \frac{16\pi}{\psi^2} \Phi\Pi
& = &  0
\label{eq:ss_krr}
\\
\frac{\left[ \left( r\psi \right)^2 \alpha' \right]'}{\left(r\psi\right)^2}
         - \left[ 16 \pi \Pi^2 + \frac{3}{2} \psi^4 \lb K^r_r \rb^2 \right] \alpha
& = &
0
\label{eq:ss_alpha}
\eeq
\beq
\left( \frac{\beta}{r} \right)'  & = & \frac{3 \alpha K^r_r }{2r}
\label{eq:ss_beta}
\\
\dot \Phi & = & \beta \Phi + \frac{\alpha}{\psi^2} \Pi
\label{eq:ss_ph}
\\
\dot \Phi & = & \left( \beta \Phi + \frac{\alpha}{\psi^2} \Pi \right)'
\label{eq:ss_phi}
\\
\dot \Pi & = & \frac{1}{r^2 \psi^4} \left[ r^2 \psi^4 \left(\beta \Pi + \frac{\alpha}{\psi^2} \Phi\right)\right]' \nonumber \\
         &   & - \left[ \alpha K^r_r + 2 \beta \frac{\left(r\psi^2\right)'}{r\psi^2}\right] \Pi.
\label{eq:ss_pi}
\eeq
Here dot and prime denote derivatives with respect to
$t$ and $r$, respectively.

The evolution equations~(\ref{eq:ss_ph}-\ref{eq:ss_pi}) are
discretized using an $O(h^2)$
Crank-Nicholson scheme. 
Eqs.~(\ref{eq:ss_psi}) and~(\ref{eq:ss_krr})
are similarly discretized using $O(h^2)$ finite difference 
approximations, then solved iteratively for $\psi$ and $K^r{}_r$
at each time step.  Once $\Phi$, $\Pi$, $\psi$ and $K^r{}_r$ have 
been determined,
$\alpha$ and $\beta$ are found from $O(h^2)$ finite-difference versions of
Eqs.~(\ref{eq:ss_alpha}) and~(\ref{eq:ss_beta}).
The code is stable and second-order convergent.




\begin{thebibliography}{1}
\bibitem{maple}
               We use Waterloo Maple along with a tensor package
               written by one of the authors (MWC).
\bibitem{boundary}
               A subtle point here concerns the ``relaxation'' occurring
               on the boundaries. The finite-difference equations on the
               boundary yield algebraic equations which determine the
               given fields there ``exactly.'' The subtlety arises because
               these algebraic conditions couple neighboring boundary points
               and thus despite the fact that we solve these equations exactly,
               the residual (as computed after the entire boundary is ``relaxed'')
               will not be identically zero. We find this procedure suffices
               to keep residuals sufficiently smooth over both interior and boundary
               domains.
\end{thebibliography}
\end{document}